\documentclass[preprint,aps,prd,onecolumn,groupedaddress,showpacs]{revtex4}

\usepackage{graphicx,amsmath,amsfonts,amssymb,mathrsfs,bbm}

\allowdisplaybreaks[1]

\newcommand{\ie}{{\it i.e.}}

\newcommand{\eg}{{\it e.g.}}

\newcommand{\cf}{{\it cf.}}

\newcommand{\eq}{Eq.}
\newcommand{\eqs}{Eqs.}
\newcommand{\fig}{Fig.}
\newcommand{\figs}{Figs.}
\newcommand{\Ref}{Ref.}
\newcommand{\Refs}{Refs.}
\newcommand{\Sec}{Sec.}

\newcommand{\App}{App.}
\renewcommand{\d}{\mathrm{d}}

\begin{document}

\title{Extrinsic CPT Violation in Neutrino Oscillations in Matter}

\author{Magnus Jacobson}
\email{magnus@theophys.kth.se}
\affiliation{Division of Mathematical Physics, Department of
  Physics, Royal Institute of Technology (KTH) -- Stockholm Center for
  Physics, Astronomy, and Biotechnology (SCFAB), Roslagstullsbacken
  11, SE-106~91~Stockholm, Sweden}

\author{Tommy Ohlsson}
\email{tommy@theophys.kth.se}
\affiliation{Division of Mathematical Physics, Department of
  Physics, Royal Institute of Technology (KTH) -- Stockholm Center for
  Physics, Astronomy, and Biotechnology (SCFAB), Roslagstullsbacken
  11, SE-106~91~Stockholm, Sweden}

\date{\today}
         
\begin{abstract}
We investigate matter-induced (or extrinsic) CPT violation effects in
neutrino oscillations in matter. Especially, we present approximate
analytical formulas for the CPT-violating probability differences for
three flavor neutrino oscillations in matter with an arbitrary matter
density profile. Note that we assume that the CPT invariance theorem
holds, which means that the CPT violation effects arise entirely
because of the presence of matter. As special cases of matter density
profiles, we consider constant and step-function matter density
profiles, which are relevant for neutrino oscillation physics in
accelerator and reactor long baseline experiments as well as neutrino
factories. Finally, the implications of extrinsic CPT violation on neutrino
oscillations in matter for several past, present, and future long
baseline experiments are estimated.
\end{abstract}

\maketitle

\section{Introduction}

Recently, several studies on CPT violation
\cite{Coleman:1997xq,Coleman:1998ti,Barger:2000iv,Murayama:2000hm,Banuls:2001zn,Barenboim:2001ac,Barenboim:2002rv,Barenboim:2002hx,Barenboim:2002tz,Barenboim:2002ah,Pakvasa:2001yw,Xing:2001ys,Skadhauge:2001kk,Bilenky:2001ka,Ohlsson:2002ne,Bahcall:2002ia,Strumia:2002fw,Mocioiu:2002pz,DeGouvea:2002xp}
have been performed in order to incorporate the so-called LSND anomaly
\cite{Athanassopoulos:1996jb,Athanassopoulos:1998pv,Aguilar:2001ty}
within the description of standard three flavor neutrino
oscillations. However, this requires a new mass squared difference
different from the ones coming from atmospheric
\cite{Fukuda:1998mi,Fukuda:1998ah,Fukuda:2000np,Shiozawa:2002}
and solar
\cite{Fukuda:2001nj,Fukuda:2001nk,Smy:2001wf,Fukuda:2002pe,Smy:2002,Ahmad:2001an,Ahmad:2002jz,Ahmad:2002ka,Hallin:2002}
neutrinos, which means that one would need to have three mass squared
differences instead of two -- a scenario, which is not consistent with
ordinary models of three flavor neutrino oscillations. Therefore, in
most of the studies on CPT violation
\cite{Murayama:2000hm,Barenboim:2001ac,Barenboim:2002rv,Barenboim:2002hx,Barenboim:2002ah,Skadhauge:2001kk,Bilenky:2001ka,Ohlsson:2002ne,Bahcall:2002ia,Strumia:2002fw},
different mass squared differences and mixing parameters are
introduced phenomenologically by hand for neutrinos and
antineutrinos. This results, in the three 
neutrino flavor picture, in two mass squared differences and four
mixing parameters for neutrinos and the same for antineutrinos, \ie,
in total, four mass squared differences and eight mixing
parameters. Thus, it is possible to have a different mass squared
difference describing the results of the LSND experiment other than the ones
describing atmospheric and solar neutrino data. It should be noted
that the results of the LSND experiment will be further tested by the
MiniBooNE experiment \cite{miniboone}, which started running in September 2002.
Furthermore, it should be mentioned that the standard way of
incorporating the LSND data is to introduce sterile neutrinos, and
therefore, the introduction of fundamental CPT violation, sometimes
also called {\it genuine} CPT violation, serves as an alternative
description to sterile neutrinos. However, neutrino oscillations
between pure sterile flavors and active and sterile flavors have, in
principle, been excluded by the SNO experiment
\cite{Ahmad:2002jz,Ahmad:2002ka,Bahcall:2002hv}.

In CPT violation studies, the CPT invariance theorem
\cite{Luders:1954,Pauli:1955,Bell:1955}, a milestone of local quantum
field theory, obviously does not hold, and in addition, fundamental
properties such as Lorentz invariance and locality may also be violated.
However, the ${\rm SU(3) \times SU(2) \times U(1)}$ Standard Model
(SM) of elementary particle physics, for which the CPT theorem is
valid, is in very good agreement with all existing experimental
data. Therefore, fundamental CPT violation is connected to physics
beyond the SM such as string theory or models including extra
dimensions, in which CPT invariance could be violated.

The recent and the first results of the KamLAND experiment
\cite{Eguchi:2002dm}, which is a reactor long baseline neutrino oscillation
experiment measuring the $\bar\nu_e$ flux from distant nuclear
reactors in Japan and South Korea, strongly favor the large mixing angle (LMA)
solution region for solar neutrino oscillations and the solar neutrino
problem \cite{Bahcall:2002ij}. Therefore, they indicate that there is
no need for fundamental CPT violation, \ie, having different mass squared
differences for solar neutrinos and reactor antineutrinos. Thus, solar
neutrino data and KamLAND data can be simultaneously and consistently
accommodated with the same mass squared difference.

In this paper, we investigate matter-induced (or extrinsic) CPT violation
effects in neutrino oscillations in matter. In a previous paper
\cite{Akhmedov:2001kd}, the interplay between fundamental and
matter-induced T violation effects has been discussed. In the case of
CPT violation effects, there exists no fundamental (or intrinsic) CPT
violation effects if we assume that the CPT theorem holds. This means that the
matter-induced CPT violation is a pure effect of the simple fact that
ordinary matter consists of unequal numbers of particles and
antiparticles. Matter-induced CPT violation, sometimes also called
{\it fake} CPT violation, have been studied and illustrated in
some papers
\cite{Bernabeu:1999gr,Banuls:2001zn,Bernabeu:2001xn,Bernabeu:2001jb,Xing:2001ys,Bernabeu:2002tj},
in which numerical calculations of CPT-violating asymmetries between survival
probabilities for neutrinos and antineutrinos in different scenarios
of atmospheric and long-baseline neutrino oscillation experiments have
been presented. Here we will try to perform a much more systematic study.

The paper is organized as follows. In \Sec~\ref{sec:genform}, we
discuss the general formalism and properties of CPT violation in vacuum and
in matter. In particular, we derive approximate analytical formulas for
all CPT-violating probability differences for three flavor neutrino
oscillations in matter with an arbitrary matter density profile. The
derivations are performed using first order perturbation theory in the
small leptonic mixing angle $\theta_{13}$ for the neutrino and
antineutrino evolution operators as well as the fact that $\Delta
m_{21}^2 \ll \Delta m_{31}^2 \simeq \Delta m_{32}^2$, \ie, the solar
mass squared difference is some orders of magnitude smaller than the
atmospheric mass squared difference. At the end of this section, we
consider two different explicit examples of matter density profiles. These are
constant and step-function matter density profiles. In both cases, we
present the first order perturbation theory formulas for the CPT
probability differences as well as the useful corresponding low-energy region
formulas. Next, in \Sec~\ref{sec:numcal}, we discuss the implications
for long baseline neutrino oscillation experiments and potential
neutrino factory setups as well as solar and
atmospheric neutrinos. We illuminate the
discussion with several tables and plots of the CPT probability
differences. Then, in \Sec~\ref{sec:s&c}, we present a summary of the
obtained results as well as our conclusions. Finally, in
\App~\ref{app:evolop}, we give details of the general analytical
derivation of the evolution operators for neutrinos and antineutrinos.

\section{General Formalism and CPT-Violating Probability Differences}
\label{sec:genform}

\subsection{Neutrino Oscillation Transition Probabilities and CP, T,
  and CPT Violation}

Let us by $P(\nu_\alpha \to \nu_\beta$) denote the transition probability
from a neutrino flavor $\alpha$ to a neutrino flavor $\beta$, and
similarly, for antineutrino flavors. Then,
the CP, T, and CPT (-violating) probability differences are given by
\begin{eqnarray}
\Delta P_{\alpha \beta}^{\rm CP} &\equiv& P(\nu_\alpha \to \nu_\beta)
- P(\bar\nu_\alpha \to \bar\nu_\beta), \label{eq:CP}\\
\Delta P_{\alpha \beta}^{\rm T} &\equiv& P(\nu_\alpha \to \nu_\beta)
- P(\nu_\beta \to \nu_\alpha), \label{eq:T}\\
\Delta P_{\alpha \beta}^{\rm CPT} &\equiv& P(\nu_\alpha \to \nu_\beta)
- P(\bar\nu_\beta \to \bar\nu_\alpha), \label{eq:CPT}
\end{eqnarray}
where $\alpha,\beta = e,\mu,\tau,\ldots$.
The CP and T probability differences have previously been
extensively studied in the literature
\cite{Arafune:1997hd,Minakata:1998bf,Minakata:1997td,Schubert:1999ky,Dick:1999ed,Donini:1999jc,Cervera:2000kp,Minakata:2000ee,Barger:2000ax,Fishbane:2000kw,Gonzalez-Garcia:2001mp,Minakata:2001qm,Minakata:2002qe,Kimura:2002wd,Yokomakura:2002av,Ota:2002fu,Kuo:1987km,Krastev:1988yu,Toshev:1989vz,Toshev:1991ku,Arafune:1997bt,Koike:1997dh,Bilenky:1998dd,Barger:1999hi,Koike:1999hf,Yasuda:1999uv,Sato:1999wt,Koike:1999tb,Harrison:1999df,Naumov:1992rh,Naumov:1992ju,Yokomakura:2000sv,Parke:2000hu,Koike:2000jf,Miura:2001pi,Akhmedov:2001kd,Miura:2001pi2,Rubbia:2001pk,Xing:2001cx,Ota:2001cz,Ohlsson:2001ps,Bueno:2001jd,Sato:2001dt,Akhmedov:2002za,Minakata:2002qi,Huber:2002uy,Leung:2003cb}.
In this paper, we will study in detail the CPT probability
differences. Let us first dicuss some general properties of the CPT
probability differences. In general, \ie, both in vacuum and in
matter, it follows from conservation of probability that
\begin{eqnarray}
\sum_{\alpha = e,\mu,\tau,\ldots} P(\nu_\alpha \to \nu_\beta) &=& 1,
  \quad \beta = e,\mu,\tau,\ldots, \label{eq:probsum1}\\
\sum_{\beta = e,\mu,\tau,\ldots} P(\nu_\alpha \to \nu_\beta) &=& 1, \quad
  \alpha = e,\mu,\tau,\ldots. \label{eq:probsum2}\
\end{eqnarray}
In words, the sum of the transition probabilities of a given neutrino
(antineutrino) flavor into neutrinos (antineutrinos) of all possible
flavors is, of course, equal to one, \ie, the probability is conserved.
Using the definitions of the CPT probability differences,
\eqs~(\ref{eq:probsum1}) and (\ref{eq:probsum2}) can be re-written as
\begin{eqnarray}
\sum_{\alpha = e,\mu,\tau,\ldots} \Delta P_{\alpha\beta}^{\rm CPT} &=& 0,
  \quad \beta = e,\mu,\tau,\ldots, \label{eq:CPTsum1}\\
\sum_{\beta = e,\mu,\tau,\ldots} \Delta P_{\alpha\beta}^{\rm CPT} &=& 0,
  \quad \alpha = e,\mu,\tau,\ldots. \label{eq:CPTsum2}
\end{eqnarray}
Note that not all of these equations are linearly independent.
For example, for three neutrino flavors, \eqs~(\ref{eq:CPTsum1}) and
(\ref{eq:CPTsum2}) can be written as the following system of equations
\begin{eqnarray}
\Delta P_{ee}^{\rm CPT} + \Delta P_{e\mu}^{\rm CPT} + \Delta
P_{e\tau}^{\rm CPT} &=& 0, \label{eq:probsum_ex}\\
\Delta P_{\mu e}^{\rm CPT} + \Delta P_{\mu\mu}^{\rm CPT} + \Delta
P_{\mu\tau}^{\rm CPT} &=& 0, \\
\Delta P_{\tau e}^{\rm CPT} + \Delta P_{\tau\mu}^{\rm CPT} + \Delta
P_{\tau\tau}^{\rm CPT} &=& 0, \\
\Delta P_{ee}^{\rm CPT} + \Delta P_{\mu e}^{\rm CPT} + \Delta
P_{\tau e}^{\rm CPT} &=& 0, \\
\Delta P_{e\mu}^{\rm CPT} + \Delta P_{\mu\mu}^{\rm CPT} + \Delta
P_{\tau\mu}^{\rm CPT} &=& 0, \\
\Delta P_{e\tau}^{\rm CPT} + \Delta P_{\mu\tau}^{\rm CPT} + \Delta
P_{\tau\tau}^{\rm CPT} &=& 0. \label{eq:probsum_xt}
\end{eqnarray}
Hence, there are nine CPT probability differences for neutrinos and
six equations relating these CPT probability differences.
The rank of the corresponding system matrix for the above system of
equations is five, which means that only five of the six equations are
linearly independent. Thus, five out of the nine CPT probability
differences can be expressed in terms of the other four, \ie, there
are, in fact, only four CPT probability differences. Choosing, \eg,
$\Delta P_{ee}^{\rm CPT}$, $\Delta P_{e\mu}^{\rm CPT}$, $\Delta P_{\mu
  e}^{\rm CPT}$, and $\Delta P_{\mu\mu}^{\rm CPT}$ as the known CPT
probability differences, the other five can be expressed as
\begin{eqnarray}
\Delta P_{e\tau}^{\rm CPT} &=& - \Delta P_{ee}^{\rm CPT} - \Delta
P_{e\mu}^{\rm CPT},\\
\Delta P_{\mu\tau}^{\rm CPT} &=& - \Delta P_{\mu e}^{\rm CPT} - \Delta
P_{\mu\mu}^{\rm CPT},\\
\Delta P_{\tau e}^{\rm CPT} &=& - \Delta P_{ee}^{\rm CPT} - \Delta
P_{\mu e}^{\rm CPT},\\
\Delta P_{\tau\mu}^{\rm CPT} &=& - \Delta P_{e\mu}^{\rm CPT} - \Delta
P_{\mu\mu}^{\rm CPT},\\
\Delta P_{\tau\tau}^{\rm CPT} &=& \Delta P_{ee}^{\rm CPT} + \Delta
P_{e\mu}^{\rm CPT} + \Delta P_{\mu e}^{\rm CPT} + \Delta
P_{\mu\mu}^{\rm CPT}.
\end{eqnarray}
Furthermore, the CPT probability differences for neutrinos are related
to the ones for antineutrinos by
\begin{eqnarray}
\Delta P_{\alpha\beta}^{\rm CPT} &=& P(\nu_\alpha \to \nu_\beta) -
P(\bar\nu_\beta \to \bar\nu_\alpha) \nonumber\\
&=& - (P(\bar\nu_\beta \to \bar\nu_\alpha) - P(\nu_\alpha \to
\nu_\beta)) = - \Delta P_{\bar\beta\bar\alpha}^{\rm CPT}, 
\end{eqnarray}
where $\alpha,\beta = e,\mu,\tau,\ldots$. Thus, the CPT probability
differences for antineutrinos do not give any further information.

For completeness, we shall also briefly consider the case of two neutrino
flavors. In this case, we have
\begin{eqnarray}
\Delta P_{ee}^{\rm CPT} + \Delta P_{e\mu}^{\rm CPT} &=& 0, \\
\Delta P_{\mu e}^{\rm CPT} + \Delta P_{\mu\mu}^{\rm CPT} &=& 0, \\
\Delta P_{ee}^{\rm CPT} + \Delta P_{\mu e}^{\rm CPT} &=& 0, \\
\Delta P_{e\mu}^{\rm CPT} + \Delta P_{\mu\mu}^{\rm CPT} &=& 0
\end{eqnarray}
from which one immediately obtains
\begin{equation}
\Delta P_{ee}^{\rm CPT} = \Delta P_{\mu\mu}^{\rm CPT} = - \Delta
P_{e\mu}^{\rm CPT} = - \Delta P_{\mu e}^{\rm CPT}.
\end{equation}
Thus, for two neutrino flavors there is only one linearly independent
CPT probability difference, which we, \eg, can choose as $\Delta
P_{ee}^{\rm CPT}$.

Generally, for the T probability differences, we have
\cite{Krastev:1988yu,Akhmedov:2001kd}
\begin{eqnarray}
&& \Delta P_{ee}^{\rm T} = \Delta P_{\mu\mu}^{\rm T} = \Delta
  P_{\tau\tau}^{\rm T} = 0, \label{eq:DPdiag}\\
&& \Delta P_{e\mu}^{\rm T} = \Delta P_{\mu\tau}^{\rm T} = \Delta
  P_{\tau e}^{\rm T} = - \Delta P_{\mu e}^{\rm T} = - \Delta
  P_{\tau\mu}^{\rm T} = - \Delta P_{e\tau}^{\rm T}
\end{eqnarray}
for three neutrino flavors and
\begin{equation}
\Delta P_{ee}^{\rm T} = \Delta P_{e\mu}^{\rm T} = \Delta P_{\mu
  e}^{\rm T} = \Delta P_{\mu\mu}^T = 0
\end{equation}
for two neutrino flavors. Thus, in the case of three neutrino flavors,
there is only one linearly independent T probability difference,
whereas in the case of two neutrino flavors, neutrino oscillations are
T-invariant irrespective of whether they take place in vacuum or in matter.

Using the definitions~(\ref{eq:CP}) - (\ref{eq:CPT}), one immediately observes
that the CP probability differences are directly related to the T and CPT
probability differences by the following formulas
\begin{equation}
\Delta P_{\alpha\beta}^{\rm CP} + \Delta P_{\bar\alpha\bar\beta}^{\rm T}
= \Delta P_{\alpha\beta}^{\rm CPT} \quad \mbox{and} \quad \Delta
P_{\bar\alpha\bar\beta}^{\rm CP} + \Delta P_{\alpha\beta}^{\rm T} =
\Delta P_{\bar\alpha\bar\beta}^{\rm CPT}.
\end{equation}
In vacuum, where CPT invariance holds, one has $\Delta
P_{\alpha\beta}^{\rm CPT} = \Delta P_{\bar\alpha\bar\beta}^{\rm CPT} =
0$, which means that $\Delta P_{\alpha\beta}^{\rm CP} = - \Delta
P_{\bar\alpha\bar\beta}^{\rm T}$ and $\Delta
P_{\bar\alpha\bar\beta}^{\rm CP} = - \Delta P_{\alpha\beta}^{\rm
T}$. Furthermore, using again the definition~(\ref{eq:CP}), one
finds that $\Delta P_{\alpha\beta}^{\rm CP} = - \Delta
P_{\bar\alpha\bar\beta}^{\rm CP}$. Thus, $\Delta P_{\alpha\beta}^{\rm
CP} = \Delta P_{\alpha\beta}^{\rm T}$ and $\Delta
P_{\bar\alpha\bar\beta}^{\rm CP} = \Delta P_{\bar\alpha\bar\beta}^{\rm
T}$, \ie, the CP probability differences for neutrinos (antineutrinos)
are given by the corresponding T probability differences for neutrinos
(antineutrinos). However, in matter, CPT invariance is
no longer valid in general, and thus, one has $\Delta
P_{\alpha\beta}^{\rm CPT} \neq 0$, which means that we need to know
both the T and CPT probability differences in order to determine the
CP probability differences. Moreover, in vacuum, it follows in general
that $P(\nu_\alpha \to \nu_\beta) = P(\bar\nu_\beta \to
\bar\nu_\alpha)$ and in particular that $P(\nu_\alpha \to \nu_\alpha)
= P(\bar\nu_\alpha \to \bar\nu_\alpha)$, which leads to $\Delta
P_{\alpha\alpha}^{\rm CP} = 0$. Therefore, CP violation effects
cannot occur in disappearance channels ($\nu_\alpha \to \nu_\alpha$),
but only in appearance channels ($\nu_\alpha \to \nu_\beta$, where
$\alpha \neq \beta$) \cite{Dick:1999ed}, while in matter one has in general
$\Delta P_{\alpha\alpha}^{\rm CP} \neq 0$.

In the next subsection, we discuss the Hamiltonians and evolution
operators for neutrinos and antineutrinos, which we will use to
calculate the CPT probability differences.

\subsection{Hamiltonians and Evolution Operators for Neutrinos and
  Antineutrinos}

If neutrinos are massive and mixed, then the neutrino flavor fields
$\nu_\alpha$, where $\alpha = e,\mu,\tau,\ldots$, are linear
combinations of the neutrino mass eigenfields $\nu_a$, where $a =
1,2,3,\ldots$, \ie,
\begin{equation}
\nu_\alpha = \sum_{a=1}^n U_{\alpha a} \nu_a, \quad \alpha =
e,\mu,\tau,\ldots,
\end{equation}
where $n$ is the number of neutrino flavors and the $U_{\alpha a}$'s
are the matrix elements of the unitary leptonic mixing matrix
$U$.\footnote{For three neutrino flavors, we use the standard
  parameterization of the unitary leptonic mixing matrix $U =
  U(\theta_{12}, \theta_{13}, \theta_{23}, \delta_{\rm CP})$
  \cite{Hagiwara:2002fs}, 
  where $s_{ij} \equiv \sin \theta_{ij}$ and $c_{ij} \equiv \cos
  \theta_{ij}$. Here $\theta_{12}$, $\theta_{13}$, and $\theta_{23}$
  are the leptonic mixing angles and $\delta_{\rm CP}$ is the leptonic
  CP violation phase.}
Thus, we have the following relation between
the neutrino flavor and mass states \cite{Bilenky:1978nj,Bilenky:1987ty}
\begin{equation}
| \nu_\alpha \rangle = \sum_{a=1}^n U_{\alpha a}^\ast | \nu_a \rangle, \quad
  \alpha = e,\mu,\tau,\ldots,
\end{equation}
where $\nu_a$ is the $a$th neutrino mass state for a neutrino with definite
3-momentum ${\bf p}$, energy $E_a = \sqrt{m_a^2 + {\bf p}^2} \simeq p +
\tfrac{m_a^2}{2p}$ (if $m_a \ll p$), and negative helicity. Here $m_a$
is the mass of the $a$th neutrino mass eigenstate and $p \equiv | {\bf
  p} |$. Similarly, for antineutrinos, we have
\begin{equation}
| \bar\nu_\alpha \rangle = \sum_{a=1}^n U_{\alpha a} | \bar\nu_a
  \rangle, \quad \alpha = e,\mu,\tau,\ldots.
\end{equation}
In the ultra-relativistic approximation, the quantum mechanical time
evolution of the neutrino states and the neutrino oscillations are
governed by the Schr{\"o}dinger equation
\begin{equation}
{\rm i} \frac{\d}{\d t} | \nu(t) \rangle = \mathscr{H}(t) | \nu(t) \rangle,
\label{eq:Schrodinger}
\end{equation}
where $| \nu(t) \rangle$ is the neutrino vector of state and
$\mathscr{H}(t)$ is the time-dependent Hamiltonian of the
system, which is different for neutrinos and antineutrinos and its
form also depends on in which basis it is given (see \App~\ref{app:evolop} for
the different expressions of the Hamiltonian).
Hence, the neutrino evolution (\ie, the solution to the
Schr{\"o}dinger equation) is given by
\begin{equation}
| \nu(t) \rangle = {\rm e}^{-{\rm i} \int_{t_0}^t \mathscr{H}(t') \d t'} |
  \nu(t_0) \rangle,
\end{equation}
where the exponential function is time-ordered. Note that if one
assumes that neutrinos are stable and that they are not absorbed in
matter, then the Hamiltonian $\mathscr{H}(t)$ is Hermitian. This will
be assumed throughout this paper. Furthermore, it is convenient to
define the evolution operator (or the evolution matrix) $S(t,t_0)$ as
\begin{equation}
| \nu(t) \rangle = S(t,t_0) | \nu(t_0) \rangle, \quad S(t,t_0) \equiv
  {\rm e}^{-{\rm i} \int_{t_0}^t \mathscr{H}(t') \d t'},
\label{eq:Smatrix}
\end{equation}
which has the following obvious properties
\begin{eqnarray}
S(t,t_0) &=& S(t,t_1) S(t_1,t_0),\\
S(t_0,t_0) &=& \mathbbm{1},\\
S(t,t_0) S(t,t_0)^\dagger &=& \mathbbm{1}.
\end{eqnarray}
The last property is the unitarity condition, which follows directly from the
hermiticity of the Hamiltonian $\mathscr{H}(t)$.

Neutrinos are produced in weak interaction processes as flavor states
$| \nu_\alpha \rangle$, where $\alpha = e,\mu,\tau,\ldots$. Between a
source, the production point of neutrinos, and a detector, neutrinos
evolve as mass eigenstates $| \nu_a \rangle$, where $a =
1,2,3,\ldots$, \ie, states with definite mass. Thus, if at time $t =
t_0$ the neutrino vector of state is $| \nu_\alpha \rangle \equiv |
\nu_\alpha(t_0) \rangle$, then at a time $t$ we have
\begin{equation}
| \nu_\alpha(t) \rangle = \sum_{a=1}^n \left[ S(t,t_0) \right]_{aa} U_{\alpha
  a}^\ast | \nu_a \rangle.
\end{equation}

The neutrino oscillation probability amplitude from a neutrino
flavor $\alpha$ to a neutrino flavor $\beta$ is defined as
\begin{equation}
A_{\alpha\beta} \equiv \langle \nu_\beta | \nu_\alpha (t) \rangle =
\sum_{a=1}^n U_{\beta a} \left[ S(t,t_0) \right]_{aa} U_{\alpha
  a}^\ast, \quad \alpha,\beta = e,\mu,\tau,\ldots.
\end{equation}
Then, the neutrino oscillation transition probability for $\nu_\alpha
\to \nu_\beta$ is given by
\begin{equation}
P(\nu_\alpha \to \nu_\beta) \equiv | A_{\alpha\beta} |^2 =
\sum_{a=1}^n \sum_{b=1}^n U_{\alpha a}^\ast U_{\beta a} U_{\alpha b}
U_{\beta b}^\ast \left[ S(t,t_0) \right]_{aa} \left[ S(t,t_0)
  \right]^\ast_{bb},
\label{eq:prob_nu}
\end{equation}
where $\alpha,\beta = e,\mu,\tau,\ldots$.

The oscillation transition probabilities for antineutrinos are
obtained by making the replacements $U_{\alpha a} \to U_{\alpha
  a}^\ast$ and $S(t,t_0) \to \bar{S}(t,t_0)$ [\ie, $V(t) \to - V(t)$,
  where $V(t)$ is the matter potential defined in \App~\ref{app:evolop}],
which lead to
\begin{eqnarray}
P(\bar\nu_\alpha \to \bar\nu_\beta) &=& \sum_{a=1}^n \sum_{b=1}^n
U_{\alpha a} U_{\beta a}^\ast U_{\alpha b}^\ast U_{\beta b} \left[
  \bar{S}(t,t_0) \right]_{aa} \left[ \bar{S}(t,t_0) \right]^\ast_{bb}
= \{ a \leftrightarrow b \} \nonumber\\
&=& \sum_{a=1}^n \sum_{b=1}^n U_{\alpha a}^\ast U_{\beta a} U_{\alpha
  b} U_{\beta b}^\ast \left[ \bar{S}(t,t_0) \right]^\ast_{aa} \left[
  \bar{S}(t,t_0) \right]_{bb},
\label{eq:prob_antinu}
\end{eqnarray}
where $\alpha,\beta = e,\mu,\tau,\ldots$.

In the next subsection, we calculate the CPT probability differences
both in vacuum and in matter.

\subsection{CPT Probability Differences}

In vacuum, the matter potential is zero, \ie, $V(t) = 0 \; \forall t$, and
therefore, the evolution operators for neutrinos and antineutrinos are
the same, \ie, $S(t,t_0) = \bar{S}(t,t_0) = {\rm e}^{-{\rm i} H_m L}$,
where $H_m = {\rm diag}\,(E_1,E_2,\ldots,E_n)$ is the free Hamiltonian
and $L \simeq t - t_0$ is the baseline length. Note that the
Hamiltonians in vacuum for neutrinos and antineutrinos are the same,
since we have assumed the CPT theorem. Thus, using
\eqs~(\ref{eq:prob_nu}) and (\ref{eq:prob_antinu}), it directly follows that
\begin{equation}
\Delta P_{\alpha\beta}^{\rm CPT} = P(\nu_\alpha \to \nu_\beta) -
P(\bar\nu_\beta \to \bar\nu_\alpha) = 0,
\label{eq:CPTvacuum}
\end{equation}
which means that there is simply no (intrinsic) CPT
violation in neutrino oscillations in vacuum. Note that this general
result holds for any number of neutrino flavors. Furthermore, note
that even though there is no intrinsic CPT violation effects in
vacuum, there could be intrinsic CP and T violation effects induced by
a non-zero CP (or T) violation phase $\delta_{\rm CP}$, which could, if
sizeable enough, be measured by very long baseline neutrino oscillation
experiments in the future \cite{Lindner:2002}.

In matter, the situation is slightly more complicated than in
vacuum. However, the technique is the same, \ie, the
extrinsic CPT probability differences are given by
differences of different matrix elements of the evolution operators
for neutrinos and antineutrinos.

The probability amplitude of neutrino flavor transitions are the matrix
elements of the evolution operators:
\begin{eqnarray}
  A \left( \nu_{\alpha} \rightarrow \nu_{\beta} \right) 
  &=& \left[ S(t, t_0) \right]_{\beta\alpha} = \left[ S_f(t, t_0)
  \right]_{\beta\alpha},\\
  A \left( \bar\nu_{\alpha} \rightarrow \bar\nu_{\beta} \right) 
  &=& \left[ \bar S(t, t_0) \right]_{\beta\alpha} = \left[ \bar S_f(t,
  t_0) \right]_{\beta\alpha}.
\end{eqnarray}
Thus, we have the extrinsic CPT probability differences
\begin{equation}
\Delta P_{\alpha\beta}^{CPT} = \left|\left[ S_f(t, t_0)
  \right]_{\beta\alpha}\right|^2 - \left|\left[ \bar S_f(t, t_0)
  \right]_{\alpha\beta}\right|^2.
\end{equation}

In the case of three neutrino flavors with the evolution operators for
neutrinos and antineutrinos as in \eqs~(\ref{eq:Sf}) and (\ref{eq:antiSf}),
respectively, the different $\Delta P_{\alpha\beta}^{\rm CPT}$'s are now
easily found, but the expressions are quite unwieldy. The CPT
probability difference $\Delta P_{ee}^{\rm CPT}$ to first order in
perturbation theory is found to be given by (see \App~\ref{app:evolop}
for definitions of different quantities)
\begin{eqnarray}
    \Delta P_{ee}^{\rm CPT} &\simeq& |S_{f,11}|^2 - |\bar S_{f,11}|^2 
    = |\alpha|^2 - |\bar\alpha|^2 = |\bar\beta|^2 - |\beta|^2 \nonumber\\
    &=& \cos^2\Omega + \frac{\sin^2\Omega}{4\Omega^2} \left( \cos
      2\theta_{12} \delta (t - t_0) - \int_{t_0}^t V(t') \d t'
    \right)^2 \nonumber\\
    &-& \cos^2\bar\Omega - \frac{\sin^2 \bar\Omega}{4\bar
      \Omega^2} \left( \cos 2\theta_{12} \delta (t - t_0) + \int_{t_0}^t
      V(t') \d t' \right)^2 \nonumber\\
    &=& \frac{1}{4} \left( \frac{\sin^2 \bar\Omega}{\bar\Omega^2} -
    \frac{\sin^2 \Omega}{\Omega^2} \right) \sin^2 2\theta_{12}
    \delta^2 (t-t_0)^2,
\end{eqnarray}
which is equal to zero in vacuum, in which $V(t) = 0 \; \forall t$. Note
that in the case of T violation all diagonal elements, \ie, $\Delta
P_{\alpha\alpha}^{\rm T}$, where $\alpha = e,\mu,\tau$, are trivially
equal to zero [\cf, \eq~(\ref{eq:DPdiag})]. This is
obviously not the case for CPT violation if matter is
present. Similarly, we find
\begin{eqnarray}
\Delta P_{e\mu}^{\rm CPT} &\simeq& |S_{f,21}|^2 - |\bar S_{f,12}|^2 =
    |c_{23}\beta^\ast + {\rm i} s_{23}fC|^2 - |c_{23}\bar\beta -
      {\rm i} s_{23}\bar f \bar A|^2 \nonumber\\
    &=& c_{23}^2\left(|\beta|^2 -|\bar\beta|^2\right)
    + s_{23}^2\left(|C|^2 - |\bar A|^2 \right) \nonumber\\
    &+& {\rm i} s_{23} c_{23} \left(\beta fC - \beta^\ast f^\ast C^\ast +
      \bar\beta^\ast \bar f \bar A - \bar\beta\bar f^\ast \bar A^\ast
    \right),\\
\Delta P_{e\tau}^{\rm CPT} &\simeq& |S_{f,31}|^2 - |\bar S_{f,13}|^2 =
    |s_{23}\beta^\ast - {\rm i} c_{23}fC|^2 - |-s_{23}\bar\beta
      - {\rm i} c_{23} \bar f \bar A|^2 \nonumber\\
    &=& c_{23}^2 \left(|C|^2 - |\bar A|^2\right) + s_{23}^2
    \left(|\beta|^2 - |\bar\beta|^2 \right) \nonumber\\
    &-& {\rm i} s_{23} c_{23} \left(\beta fC - \beta^\ast f^\ast C^\ast +
      \bar\beta^\ast \bar f \bar A - \bar\beta\bar f^\ast \bar
      A^\ast \right),\\
\Delta P_{\mu e}^{\rm CPT} &\simeq& |S_{f,12}|^2 - |\bar S_{f,21}|^2 =
    |c_{23}\beta - {\rm i} s_{23}fA|^2 - |c_{23}\bar\beta^\ast +
      {\rm i} s_{23}\bar f \bar C|^2 \nonumber\\
    &=& c_{23}^2\left(|\beta|^2 -|\bar\beta|^2\right)
    + s_{23}^2\left(|A|^2 - |\bar C|^2 \right) \nonumber\\
    &+& {\rm i} s_{23} c_{23} \left(\beta f^\ast A^\ast - \beta^\ast fA -
      \bar\beta\bar f \bar C + \bar\beta^\ast \bar f^\ast \bar C^\ast
    \right),\\
\Delta P_{\tau e}^{\rm CPT} &\simeq& |S_{f,13}|^2 - |\bar S_{f,31}|^2 =
    |-s_{23}\beta - {\rm i} c_{23}fA|^2 - |s_{23}\bar\beta^\ast
      - {\rm i} c_{23} \bar f \bar C|^2 \nonumber\\
    &=& c_{23}^2 \left(|A|^2 - |\bar C|^2\right) + s_{23}^2
    \left(|\beta|^2 - |\bar\beta|^2 \right) \nonumber\\
    &-& {\rm i} s_{23} c_{23} \left(\beta f^\ast A^\ast - \beta^\ast fA -
      \bar\beta\bar f\bar C + \bar\beta^\ast \bar f^\ast \bar
      C^\ast \right),\\
\Delta P_{\mu\mu}^{\rm CPT} &\simeq& |S_{f,22}|^2 - |\bar S_{f,22}|^2
    = |c_{23}^2\alpha^\ast + s_{23}^2f - {\rm i} s_{23}c_{23}f(B + D)|^2
    \nonumber\\ 
    &-& |c_{23}^2\bar\alpha^\ast + s_{23}^2\bar f -
      {\rm i} s_{23}c_{23}\bar f( \bar B + \bar D)|^2 \nonumber\\
    &=& c_{23}^4 \left(|\alpha|^2 - |\bar\alpha|^2\right) - {\rm i}
    s_{23} c_{23}^3 \left(\alpha fB - \alpha^\ast f^\ast B^\ast +
    \alpha fD - \alpha^\ast f^\ast D^\ast
      \vphantom{\bar B^\ast}\right. \nonumber\\
    && \left.\vphantom{\bar B^\ast} -
      \bar\alpha\bar f\bar B + \bar\alpha^\ast \bar f^\ast \bar B^\ast -
      \bar\alpha\bar f\bar D + \bar\alpha^\ast \bar f^\ast \bar
      D^\ast \right) \nonumber\\
    && + s_{23}^2 c_{23}^2 \left(\alpha f + \alpha^\ast f^\ast + |B|^2
      + |D|^2 + BD^\ast + B^\ast D \vphantom{\bar B^\ast}\right. \nonumber\\
    && \left.\vphantom{\bar B^\ast} - \bar\alpha\bar f -
      \bar\alpha^\ast \bar f^\ast - |\bar B|^2 - |\bar D|^2 -
      \bar B\bar D^\ast - \bar B^\ast \bar D \right) \nonumber\\
    && - {\rm i} s_{23}^3 c_{23} \left(B - B^\ast + D - D^\ast - \bar B +
      \bar B^\ast - \bar D + \bar D^\ast \right),\\
\Delta P_{\mu \tau}^{\rm CPT} &\simeq& |S_{f,32}|^2 - |\bar S_{f,23}|^2
    = |-s_{23}c_{23} \left(\alpha^\ast - f\right) + {\rm i} f \left(
        s_{23}^2 B - c_{23}^2 D \right)|^2 \nonumber\\
    &-& |-s_{23}c_{23} \left( \bar\alpha^\ast - \bar f \right)
      - {\rm i} \bar f \left( c_{23}^2\bar B - s_{23}^2\bar D
    \right)|^2 \nonumber\\ 
    &=& c_{23}^4 \left(|D|^2 - |\bar B|^2\right) + {\rm i}
    s_{23} c_{23}^3 \left(\alpha fD - \alpha^\ast f^\ast D^\ast - D + D^\ast
      \vphantom{\bar B^\ast}\right. \nonumber\\
    && \left.\vphantom{\bar B^\ast}- \bar\alpha \bar f \bar B +
      \bar\alpha^\ast \bar f^\ast \bar B^\ast + \bar B - \bar B^\ast \right) +
    s_{23}^2 c_{23}^2 \left( |\alpha|^2 - \alpha f - \alpha^\ast f^\ast
      \vphantom{|\alpha|^2}\right. \nonumber\\
    && \vphantom{|\alpha|^2}\left. - BD^\ast - B^\ast D -
      |\bar\alpha|^2 + \bar\alpha\bar f + \bar\alpha^\ast \bar
      f^\ast + \bar B\bar D^\ast + \bar B^\ast \bar D \right) \nonumber\\
    && - {\rm i} s_{23}^3 c_{23} \left( \alpha fB - \alpha^\ast f^\ast
    B^\ast - B + B^\ast - \bar\alpha\bar f\bar D + \bar\alpha^\ast
    \bar f^\ast \bar D^\ast \vphantom{\bar D^\ast}\right. \nonumber\\
    && \vphantom{\bar D^\ast}\left. + \bar D - \bar D^\ast \right) +
    s_{23}^4 \left( |B|^2 - |\bar D|^2 \right),\\
\Delta P_{\tau\mu}^{\rm CPT} &\simeq& |S_{f,23}|^2 - |\bar S_{f,32}|^2
    = |-s_{23}c_{23} \left(\alpha^\ast - f\right) - {\rm i} f
      \left(c_{23}^2B - s_{23}^2 D \right)|^2 \nonumber\\
    &-& |-s_{23}c_{23} \left(\bar\alpha^\ast - \bar f\right) +
      {\rm i} \bar f \left(s_{23}^2\bar B - c_{23}^2\bar D \right)|^2
    \nonumber\\
    &=& c_{23}^4 \left(|B|^2 - |\bar D|^2\right) + {\rm i}
    s_{23} c_{23}^3 \left(\alpha fB - \alpha^\ast f^\ast B^\ast - B + B^\ast
      \vphantom{\bar D^\ast}\right. \nonumber\\
    && \left.\vphantom{\bar D^\ast}- \bar\alpha \bar f \bar D +
      \bar\alpha^\ast \bar f^\ast \bar D^\ast + \bar D - \bar D^\ast \right) +
    s_{23}^2 c_{23}^2 \left( |\alpha|^2 - \alpha f - \alpha^\ast f^\ast
      \vphantom{|\alpha|^2}\right. \nonumber\\
    && \vphantom{|\alpha|^2}\left. - BD^\ast - B^\ast D -
      |\bar\alpha|^2 + \bar\alpha\bar f + \bar\alpha^\ast \bar
      f^\ast + \bar B\bar D^\ast + \bar B^\ast \bar D \right) \nonumber\\
    && - {\rm i} s_{23}^3 c_{23} \left( \alpha fD - \alpha^\ast f^\ast
    D^\ast - D + D^\ast - \bar\alpha\bar f\bar B + \bar\alpha^\ast
    \bar f^\ast \bar B^\ast \vphantom{\bar B^\ast}\right. \nonumber\\
    && \left.\vphantom{\bar B^\ast} + \bar B - \bar B^\ast \right) +
    s_{23}^4 \left( |D|^2 - |\bar B|^2 \right),\\
\Delta P_{\tau\tau}^{\rm CPT} &\simeq& |S_{f,33}|^2 - |\bar S_{f,33}|^2
    = |s_{23}^2\alpha^\ast + c_{23}^2f + {\rm i} s_{23}c_{23}f
      \left(B + D\right)|^2 \nonumber\\
    &-& |s_{23}^2\bar\alpha^\ast + c_{23}^2\bar f + {\rm i}
      s_{23}c_{23}\bar f \left(\bar B + \bar D\right)|^2 \nonumber\\
    &=& {\rm i} s_{23} c_{23}^3 \left(B - B^\ast + D - D^\ast - \bar B
    + \bar B^\ast - \bar D + \bar D^\ast\right) \nonumber\\
    && + s_{23}^2 c_{23}^2 \left(\alpha f + \alpha^\ast f^\ast + |B|^2
      + |D|^2 + BD^\ast + B^\ast D \vphantom{\bar B^\ast}\right. \nonumber\\
    && \left.\vphantom{\bar B^\ast} - \bar\alpha\bar f -
      \bar\alpha^\ast \bar f^\ast - |\bar B|^2 - |\bar D|^2 -
      \bar B\bar D^\ast - \bar B^\ast \bar D \right) \nonumber\\
    && + {\rm i} s_{23}^3 c_{23} \left(\alpha fB - \alpha^\ast f^\ast B^\ast +
      \alpha fD - \alpha^\ast f^\ast D^\ast - \bar\alpha\bar f\bar B
      \vphantom{\bar B^\ast}\right. \nonumber\\
    && \left.\vphantom{\bar B^\ast} + \bar\alpha^\ast \bar f^\ast \bar
      B^\ast - \bar\alpha\bar f \bar D + \bar\alpha^\ast \bar f^\ast \bar
      D^\ast \right) + s_{23}^4 \left(|\alpha|^2 -
      |\bar\alpha|^2\right).
\end{eqnarray}
Note that $\Delta P_{ee}^{\rm CPT}$ is the only CPT probability
difference that is uniquely determined by the (1,2)-subsector of the
full three flavor neutrino evolution, see the explicit expressions of
the evolution operators for neutrinos and antineutrinos [\eqs~(\ref{eq:Sf}) and
(\ref{eq:antiSf})]. Thus, it is completely independent of the CP
violation phase $\delta_{\rm CP}$ \cite{Xing:2001ys} as well as the fundamental
neutrino parameters $\Delta m_{31}^2 \simeq \Delta m_{32}^2$,
$\theta_{13}$, and $\theta_{23}$.

Now, using conservation of probability, \ie,
\eqs~(\ref{eq:probsum_ex}) - (\ref{eq:probsum_xt}), we find the
relations
\begin{eqnarray}
\sum_{\alpha = e,\mu,\tau} \Delta P_{e\alpha}^{\rm CPT} &=& |C|^2 -
|\bar A|^2 = 0,\\
\sum_{\alpha = e,\mu,\tau} \Delta P_{\alpha e}^{\rm CPT} &=& |A|^2 -
|\bar C|^2 = 0,\\
\sum_{\alpha = e,\mu,\tau} \Delta P_{\mu\alpha}^{\rm CPT} + \sum_{\alpha =
  e,\mu,\tau} \Delta P_{\tau\alpha}^{\rm CPT} &=&
\sum_{\alpha = e,\mu,\tau} \Delta P_{\alpha\mu}^{\rm CPT} + \sum_{\alpha =
  e,\mu,\tau} \Delta P_{\alpha\tau}^{\rm CPT} \nonumber\\
&=& |B|^2 + |D|^2 - |\bar B|^2 - |\bar D|^2 = 0.
\end{eqnarray}
Thus, the CPT probability differences can be further
simplified and we obtain
\begin{eqnarray}
\Delta P_{ee}^{\rm CPT} &\simeq& |\bar\beta|^2 - |\beta|^2, \label{eq:DPee}\\
\Delta P_{e\mu}^{\rm CPT} &\simeq& c_{23}^2\left(|\beta|^2
    -|\bar\beta|^2\right)
    - 2 c_{23}s_{23} \Im \left(\beta fC - \bar\beta\bar f^\ast \bar A^\ast
    \right), \label{eq:DPemu}\\
\Delta P_{e\tau}^{\rm CPT} &\simeq& s_{23}^2 \left(|\beta|^2 -
    |\bar\beta|^2 \right)
    + 2 c_{23}s_{23} \Im \left(\beta fC - \bar\beta\bar f^\ast \bar A^\ast
    \right),\\
\Delta P_{\mu e}^{\rm CPT} &\simeq& c_{23}^2\left(|\beta|^2
    -|\bar\beta|^2\right)
    - 2 c_{23}s_{23} \Im \left(\beta f^\ast A^\ast - 
      \bar\beta\bar f \bar C \right),\\
\Delta P_{\tau e}^{\rm CPT} &\simeq& s_{23}^2 
    \left(|\beta|^2 - |\bar\beta|^2 \right)
    + 2 c_{23}s_{23} \Im \left(\beta f^\ast A^\ast - 
      \bar\beta\bar f \bar C \right), \label{eq:DPtaue}
\end{eqnarray}
where we have only displayed the CPT probability differences $\Delta
P_{ee}^{\rm CPT}$, $\Delta P_{e\mu}^{\rm CPT}$, $\Delta P_{e\tau}^{\rm
CPT}$, $\Delta P_{\mu e}^{\rm CPT}$, and $\Delta P_{\tau e}^{\rm
CPT}$, since the remaining ones are too lengthy expressions and not
so illuminating. In the following, we will restrict our discussion
only to those CPT probability differences displayed above. 
Furthermore, from the definition of the parameters $a$ and $b$ in
\eq~(\ref{eq:H1}), we can conclude that $|b/a| \propto
\delta^2/\Delta^2 = \left(\Delta m_{21}^2/\Delta m_{31}^2\right)^2$,
and thus, the ratio $|b/a|$ is small, since $\Delta m_{21}^2 \ll
\Delta m_{31}^2$. In \Ref~\cite{Akhmedov:2001kd}, it has been shown that
$$
|I_{\beta,t}(t,t_0)/I_{\alpha^\ast,t}(t,t_0)| \sim
|I^\ast_{\beta,t_0}(t,t_0)/I^\ast_{\alpha^\ast,t_0}(t,t_0)| \sim
\delta^2/\Delta^2,
$$
and therefore, it also holds that
$$
|\bar I_{\beta,t}(t,t_0)/\bar I_{\alpha^\ast,t}(t,t_0)| \sim
|\bar I^\ast_{\beta,t_0}(t,t_0)/\bar I^\ast_{\alpha^\ast,t_0}(t,t_0)| \sim
\delta^2/\Delta^2.
$$
Thus, the contributions of the integrals $I_{\beta,t}(t,t_0)$,
$I^\ast_{\beta,t_0}(t,t_0)$, $\bar I_{\beta,t}(t,t_0)$, and $\bar
I^\ast_{\beta,t_0}(t,t_0)$ are suppressed by a factor of
$\delta^2/\Delta^2$ in \eqs~(\ref{eq:DPemu}) -
(\ref{eq:DPtaue}). Using this to reduce the arguments of the
imaginary parts in \eqs~(\ref{eq:DPemu}) - (\ref{eq:DPtaue}) further, we
obtain the following
\begin{eqnarray}
\beta f C - \bar\beta \bar f^\ast \bar A^\ast &\simeq& \beta f a^\ast
I^\ast_{\alpha^\ast,t_0} - \bar\beta \bar f^\ast a \bar
I^\ast_{\alpha^\ast,t}, \\
\beta f^\ast A^\ast - \bar\beta \bar f \bar C &\simeq& \beta f^\ast a^\ast
I^\ast_{\alpha^\ast,t} - \bar\beta \bar f a \bar I^\ast_{\alpha^\ast,t_0}.
\end{eqnarray}

\subsection{Examples of Matter Density Profiles}

We have now derived the general analytical expressions for the CPT violation
probability differences.
Next, we will calculate some of the CPT violation probability differences for
some specific examples of matter density profiles. This will be done
for constant matter density and step-function matter density.

\subsubsection{Constant Matter Density Profiles}

The simplest example of a matter density profile (except for vacuum)
is the one of constant matter density or constant electron
density. In this case, the matter potential is given by $V(t) = V =
\text{const.} \; \forall t$. Furthermore, if the distance between
source and detector (\ie, the neutrino propagation path length or
baseline length) is $L$ and the neutrino energy is $E_\nu$, then we can
define the following useful quantities
\begin{eqnarray}
\omega &\equiv& \frac{\delta}{2} \sqrt{\left[ \cos 2\theta_{12} -
\frac{V}{\delta} \right]^2 + \sin^2 2\theta_{12}},\\
\bar\omega &\equiv& \frac{\delta}{2} \sqrt{\left[ \cos 2\theta_{12} +
\frac{V}{\delta} \right]^2 + \sin^2 2\theta_{12}},\\
\tilde\Delta &=& \Delta - \frac{1}{2} (V + \delta) = \frac{\delta}{2}
\left( 2 \frac{\Delta}{\delta} - 1 - \frac{V}{\delta} \right),\\
\bar{\tilde\Delta} &=& \Delta - \frac{1}{2} (- V + \delta) = \frac{\delta}{2}
\left( 2 \frac{\Delta}{\delta} - 1 + \frac{V}{\delta} \right),\\
\theta_m &\equiv& \frac{1}{2} \arccos \left( \frac{\delta \cos
  2\theta_{12} - V}{2\omega} \right),\\
\bar\theta_m &\equiv& \frac{1}{2} \arccos \left( \frac{\delta \cos
  2\theta_{12} + V}{2\bar\omega} \right),
\end{eqnarray}
where $\delta \equiv \frac{\Delta m_{21}^2}{2 E_\nu}$, $\Delta \equiv
\frac{\Delta m_{31}^2}{2 E_\nu} \simeq \frac{\Delta m_{32}^2}{2
E_\nu}$, and $\theta_{12}$ is the solar mixing angle. 
Then, we have (see \App~\ref{app:evolop})
\begin{eqnarray}
\alpha(t,0) &=& \cos \omega t + {\rm i} \cos 2\theta_m \sin \omega t,
\label{eq:alphat0}\\
\bar\alpha(t,0) &=& \cos \bar\omega t + {\rm i} \cos 2\bar\theta_m
\sin \bar\omega t,\\
\beta(t,0) &=& - {\rm i} \sin 2\theta_m \sin \omega t,\\
\bar\beta(t,0) &=& - {\rm i} \sin 2\bar\theta_m \sin \bar\omega t,\\
f(t,0) &=& {\rm e}^{- {\rm i} \tilde\Delta t},\\
\bar f(t,0) &=& {\rm e}^{- {\rm i} \bar{\tilde\Delta} t}, \label{eq:barft0}
\end{eqnarray}
where $0 \leq t \leq L$, which yield
\begin{eqnarray}
|\beta|^2 - |\bar\beta|^2 &=& \sin^2 2\theta_m s^2 - \sin^2
 2\bar\theta_m \bar s^2 \nonumber\\
&=& s_{12}^2 c_{12}^2 \delta^2
 \left( \frac{s^2}{\omega^2} - \frac{\bar s^2}{\bar\omega^2} \right),
 \label{eq:b2b2}\\
\Im \left(\beta fC - \bar\beta\bar f^\ast \bar A^\ast \right) &\simeq&
 s_{12} c_{12} s_{13} \delta \left( \Delta - s_{12}^2 \delta \right)
 \bigg\{ \left( \frac{\bar s^2}{\bar\omega^2} - \frac{s^2}{\omega^2}
 \right) \cos \delta_{\rm CP} \nonumber\\
&+& \left( \Delta - c_{12}^2 \delta
 \right) \left[ \frac{(\bar{\tilde \Delta} \bar s - \bar\omega \sin
 \bar{\tilde \Delta} L)\bar
 s}{\bar\omega^2 (\bar\omega^2 - \bar{\tilde \Delta}^2)} -
 \frac{(\tilde \Delta s - \omega \sin \tilde \Delta L) s}{\omega^2
 (\omega^2 - \tilde \Delta^2)} \right] \cos \delta_{\rm CP} \nonumber\\
&+& \left( \Delta - c_{12}^2 \delta \right) \left[ \frac{(\cos
 \bar{\tilde \Delta} L - \bar c) \bar s}{\bar\omega (\bar\omega^2 -
 \bar{\tilde \Delta}^2)} - \frac{(\cos \tilde \Delta L - c) s}{\omega
 (\omega^2 - \tilde \Delta^2)} \right] \sin \delta_{\rm CP} \bigg\},
 \label{eq:im1}\\
\Im \left(\beta f^\ast A^\ast - \bar\beta\bar f \bar C \right) &\simeq&
s_{12} c_{12} s_{13} \delta \left( \Delta - s_{12}^2 \delta \right)
 \bigg\{ \left( \frac{\bar s^2}{\bar\omega^2} - \frac{s^2}{\omega^2}
 \right) \cos \delta_{\rm CP} \nonumber\\
&+& \left( \Delta - c_{12}^2 \delta
 \right) \left[ \frac{(\bar{\tilde \Delta} \bar s - \bar\omega \sin
 \bar{\tilde \Delta} L)\bar
 s}{\bar\omega^2 (\bar\omega^2 - \bar{\tilde \Delta}^2)} -
 \frac{(\tilde \Delta s - \omega \sin \tilde \Delta L) s}{\omega^2
 (\omega^2 - \tilde \Delta^2)} \right] \cos \delta_{\rm CP} \nonumber\\
&-& \left( \Delta - c_{12}^2 \delta \right) \left[ \frac{(\cos
 \bar{\tilde \Delta} L - \bar c) \bar s}{\bar\omega (\bar\omega^2 -
 \bar{\tilde \Delta}^2)} - \frac{(\cos \tilde \Delta L - c) s}{\omega
 (\omega^2 - \tilde \Delta^2)} \right] \sin \delta_{\rm CP}
 \bigg\}, \label{eq:im2}
\end{eqnarray}
where $s \equiv \sin \omega L$, $\bar s \equiv \sin \bar\omega L$,
$c \equiv \cos \omega L$, and $\bar c \equiv \cos \bar\omega L$.
Note that the only difference between the imaginary parts in
\eqs~(\ref{eq:im1}) and (\ref{eq:im2}) is the signs in front of the
$\sin \delta_{\rm CP}$ terms, \ie, applying the replacement
$\delta_{\rm CP} \to - \delta_{\rm CP}$, one comes from $\Im(\beta f C
- \bar\beta \bar f^\ast \bar A^\ast)$ to $\Im(\beta f^\ast A^\ast -
\bar\beta \bar f \bar C)$, and vice versa.
Thus, inserting \eqs~(\ref{eq:b2b2}) - (\ref{eq:im2}) into
\eqs~(\ref{eq:DPee}) - (\ref{eq:DPtaue}), we obtain the CPT
probability differences in matter of constant density as
\begin{eqnarray}
\Delta P_{ee}^{\rm CPT} &\simeq& - s_{12}^2 c_{12}^2 \delta^2
 \left( \frac{\sin^2 \omega  L}{\omega^2} - \frac{\sin^2 \bar\omega
 L}{\bar\omega^2} \right), \label{eq:Pee}\\
\Delta P_{e\mu}^{\rm CPT} &\simeq& s_{12}^2 c_{12}^2 c_{23}^2 \delta^2
 \left( \frac{\sin^2 \omega  L}{\omega^2} - \frac{\sin^2 \bar\omega
 L}{\bar\omega^2} \right) - 2 s_{12} c_{12} s_{13} s_{23} c_{23} \delta \left(
 \Delta - s_{12}^2 \delta \right) \nonumber\\
&\times& \bigg\{ \left( \frac{\sin^2 \bar\omega L}{\bar\omega^2} - \frac{\sin^2
 \omega L}{\omega^2} \right) \cos \delta_{\rm CP} \nonumber\\
&+& \left( \Delta - c_{12}^2 \delta
 \right) \left[ \frac{(\bar{\tilde \Delta} \bar s - \bar\omega \sin
 \bar{\tilde \Delta} L)\bar
 s}{\bar\omega^2 (\bar\omega^2 - \bar{\tilde \Delta}^2)} -
 \frac{(\tilde \Delta s - \omega \sin \tilde \Delta L) s}{\omega^2
 (\omega^2 - \tilde \Delta^2)} \right] \cos \delta_{\rm CP} \nonumber\\
&+& \left( \Delta - c_{12}^2 \delta \right) \left[ \frac{(\cos
 \bar{\tilde \Delta} L - \bar c) \bar s}{\bar\omega (\bar\omega^2 -
 \bar{\tilde \Delta}^2)} - \frac{(\cos \tilde \Delta L - c) s}{\omega
 (\omega^2 - \tilde \Delta^2)} \right] \sin \delta_{\rm CP} \bigg\},
 \label{eq:Pem}\\
\Delta P_{e\tau}^{\rm CPT} &\simeq& s_{12}^2 c_{12}^2 s_{23}^2 \delta^2
 \left( \frac{\sin^2 \omega  L}{\omega^2} - \frac{\sin^2 \bar\omega
 L}{\bar\omega^2} \right) + 2 s_{12} c_{12} s_{13} s_{23} c_{23} \delta \left(
 \Delta - s_{12}^2 \delta \right) \nonumber\\
&\times& \bigg\{ \left( \frac{\sin^2 \bar\omega L}{\bar\omega^2} - \frac{\sin^2
 \omega L}{\omega^2} \right) \cos \delta_{\rm CP} \nonumber\\
&+& \left( \Delta - c_{12}^2 \delta
 \right) \left[ \frac{(\bar{\tilde \Delta} \bar s - \bar\omega \sin
 \bar{\tilde \Delta} L)\bar
 s}{\bar\omega^2 (\bar\omega^2 - \bar{\tilde \Delta}^2)} -
 \frac{(\tilde \Delta s - \omega \sin \tilde \Delta L) s}{\omega^2
 (\omega^2 - \tilde \Delta^2)} \right] \cos \delta_{\rm CP} \nonumber\\
&+& \left( \Delta - c_{12}^2 \delta \right) \left[ \frac{(\cos
 \bar{\tilde \Delta} L - \bar c) \bar s}{\bar\omega (\bar\omega^2 -
 \bar{\tilde \Delta}^2)} - \frac{(\cos \tilde \Delta L - c) s}{\omega
 (\omega^2 - \tilde \Delta^2)} \right] \sin \delta_{\rm CP} \bigg\},\\
\Delta P_{\mu e}^{\rm CPT} &\simeq& s_{12}^2 c_{12}^2 c_{23}^2 \delta^2
 \left( \frac{\sin^2 \omega  L}{\omega^2} - \frac{\sin^2 \bar\omega
 L}{\bar\omega^2} \right) - 2 s_{12} c_{12} s_{13} s_{23} c_{23} \delta \left(
 \Delta - s_{12}^2 \delta \right) \nonumber\\
&\times& \bigg\{ \left( \frac{\sin^2 \bar\omega L}{\bar\omega^2} - \frac{\sin^2
 \omega L}{\omega^2} \right) \cos \delta_{\rm CP} \nonumber\\
&+& \left( \Delta - c_{12}^2 \delta
 \right) \left[ \frac{(\bar{\tilde \Delta} \bar s - \bar\omega \sin
 \bar{\tilde \Delta} L)\bar
 s}{\bar\omega^2 (\bar\omega^2 - \bar{\tilde \Delta}^2)} -
 \frac{(\tilde \Delta s - \omega \sin \tilde \Delta L) s}{\omega^2
 (\omega^2 - \tilde \Delta^2)} \right] \cos \delta_{\rm CP} \nonumber\\
&-& \left( \Delta - c_{12}^2 \delta \right) \left[ \frac{(\cos
 \bar{\tilde \Delta} L - \bar c) \bar s}{\bar\omega (\bar\omega^2 -
 \bar{\tilde \Delta}^2)} - \frac{(\cos \tilde \Delta L - c) s}{\omega
 (\omega^2 - \tilde \Delta^2)} \right] \sin \delta_{\rm CP} \bigg\},
 \label{eq:Pme}\\
\Delta P_{\tau e}^{\rm CPT} &\simeq& s_{12}^2 c_{12}^2 s_{23}^2 \delta^2
 \left( \frac{\sin^2 \omega  L}{\omega^2} - \frac{\sin^2 \bar\omega
 L}{\bar\omega^2} \right) + 2 s_{12} c_{12} s_{13} s_{23} c_{23} \delta \left(
 \Delta - s_{12}^2 \delta \right) \nonumber\\
&\times& \bigg\{ \left( \frac{\sin^2 \bar\omega L}{\bar\omega^2} - \frac{\sin^2
 \omega L}{\omega^2} \right) \cos \delta_{\rm CP} \nonumber\\
&+& \left( \Delta - c_{12}^2 \delta
 \right) \left[ \frac{(\bar{\tilde \Delta} \bar s - \bar\omega \sin
 \bar{\tilde \Delta} L)\bar
 s}{\bar\omega^2 (\bar\omega^2 - \bar{\tilde \Delta}^2)} -
 \frac{(\tilde \Delta s - \omega \sin \tilde \Delta L) s}{\omega^2
 (\omega^2 - \tilde \Delta^2)} \right] \cos \delta_{\rm CP} \nonumber\\
&-& \left( \Delta - c_{12}^2 \delta \right) \left[ \frac{(\cos
 \bar{\tilde \Delta} L - \bar c) \bar s}{\bar\omega (\bar\omega^2 -
 \bar{\tilde \Delta}^2)} - \frac{(\cos \tilde \Delta L - c) s}{\omega
 (\omega^2 - \tilde \Delta^2)} \right] \sin \delta_{\rm CP}
 \bigg\}. \label{eq:Ptaue}
\end{eqnarray}
It is again interesting to observe that the CPT probability difference
$\Delta P_{ee}^{\rm CPT}$ contains only a constant term in the mixing
parameter $\delta_{\rm CP}$, \ie, it is independent of the CP
violation phase $\delta_{\rm CP}$, whereas the other CPT probability
differences contain such terms, but in addition also $\sin
\delta_{\rm CP}$ and $\cos \delta_{\rm CP}$ terms
(in the case of CP violation, see, \eg, \Ref~\cite{Koike:2000jf}).
Naively, one would not expect any $\sin \delta_{\rm CP}$ terms in the
CPT probability 
differences, since they do not arise in the general case of the T probability
difference as an effect of the presence of matter, but are there
because of the fundamental T violation that is caused by the CP
violation phase $\delta_{\rm CP}$ \cite{Akhmedov:2001kd}. However,
since constant matter density profiles are symmetric with respect to
the baseline length $L$, the T violation probability difference is anyway
actually equal to zero in these cases. Furthermore, we note that if
one makes the replacement $\delta_{\rm CP} \to -\delta_{\rm CP}$, then
$\Delta P_{e\mu}^{\rm CPT} \to \Delta P_{\mu e}^{\rm CPT}$ and $\Delta
P_{e\tau}^{\rm CPT} \to \Delta P_{\tau e}^{\rm CPT}$ and in the case
that $\delta_{\rm CP} = 0$ one has $\Delta P_{e\mu}^{\rm CPT} =
\Delta P_{\mu e}^{\rm CPT}$ and $\Delta P_{e\tau}^{\rm CPT} = \Delta
P_{\tau e}^{\rm CPT}$.
Moreover, in the case of degenerate neutrino masses $m_1 = m_2$ or for
extremely high neutrino energies, $E_\nu \to \infty$, the quantity $\delta =
\frac{\Delta m_{21}^2}{2E_\nu}$ goes to zero and so do $\beta$ and
$\bar\beta$ (see the second point in the discussion at the end of
\App~\ref{app:evolop} about the relation between $\Omega$ and
$\bar\Omega$), which in turn means that the CPT probability
differences in \eqs~(\ref{eq:DPee}) - (\ref{eq:DPtaue}) as well as in
\eqs~(\ref{eq:Pee}) - (\ref{eq:Ptaue}) will vanish, \ie, $\Delta
P_{\alpha\beta}^{\rm CPT} \to 0$ when $\delta \to 0$. This can be
understood as follows. In the case when $\Delta m_{21}^2 \ll \Delta
m_{31}^2$ (\ie, $\delta \ll \Delta$) or in the limit $\delta \to 0$, we
have that the neutrino mass hierarchy parameter $\xi \equiv
\frac{\Delta m_{21}^2}{\Delta m_{31}^2} = \frac{\delta}{\Delta}$ also
goes to zero. If $\xi \to 0$, then $P_{ee} \to 1 - \sin^2 2\theta_{13}
\frac{\sin^2 C_{13} \Delta L}{C_{13}^2}$, where $C_{13} \equiv
\sqrt{\sin^2 2\theta_{13} + (\frac{2}{\Delta} V - \cos
2\theta_{13})^2}$. Now, since we have only calculated the CPT
probability differences to first order in perturbation theory in the
small leptonic mixing angle $\theta_{13}$ (see \App~\ref{app:evolop}),
we have that $P_{ee} \to 1$ when $\xi \to 0$. Using $P_{ee} = 1$
together with the unitarity conditions~(\ref{eq:probsum1}) and
(\ref{eq:probsum2}), we find that $P_{\mu\mu} = P_{\tau\tau} = 1$ and
$P_{e\mu} = P_{\mu e} = P_{\mu\tau} = P_{\tau\mu} = 0$, which means
that neutrino oscillations will not occur in this limit. A similar
argument applies for the case of antineutrinos. Thus, the CPT
probability differences $\Delta P_{\alpha\beta}^{\rm CPT} \to 0$ up to
first order in perturbation theory in $\theta_{13}$ when $\delta \to
0$ (\ie, when $\delta$ is completely negligible compared with
$\Delta$). Therefore, there are no extrinsic CPT violation effects up
to first order in $\theta_{13}$ when $\delta \to 0$.

In the low-energy region $V \lesssim \delta \ll \Delta$, we find after
some tedious calculations that
\begin{eqnarray}
\Delta P_{ee}^{\rm CPT} &\simeq& 8 s_{12}^2 c_{12}^2 \cos 2\theta_{12}
\left( \delta L \cos\frac{\delta L}{2} - 2 \sin\frac{\delta L}{2}
\right) \sin\frac{\delta L}{2} \frac{V}{\delta} \nonumber\\
&+& \mathcal{O}\left((V/\delta)^3\right), \label{eq:DPeeCPTlow}\\
\Delta P_{e\mu}^{\rm CPT} &\simeq& -8 s_{12}^2 c_{12}^2 c_{23}^2 \cos
2\theta_{12} \left( \delta L \cos\frac{\delta L}{2} - 2 \sin\frac{\delta L}{2}
\right) \sin\frac{\delta L}{2} \frac{V}{\delta} \nonumber\\
&-& 16 s_{12} c_{12}^3 s_{13} s_{23} c_{23} \cos \delta_{\rm CP} \cos
2\theta_{12} \left( \delta L \cos\frac{\delta L}{2} - 2 \sin\frac{\delta
  L}{2} \right) \sin\frac{\delta L}{2} \frac{V}{\delta} \nonumber\\
&-& 16 s_{12} c_{12} s_{13} s_{23} c_{23} \sin \delta_{\rm CP} 
\Bigg\{ \cos 2\theta_{12} \bigg[ \delta L \cos \delta L - \cos \Delta
  L \nonumber\\
&\times& \left( \delta L \cos\frac{\delta L}{2} - 2 \sin\frac{\delta L}{2}
\right) - \sin \delta L \bigg] + \delta L \sin\frac{\delta L}{2} \sin
\Delta L \Bigg\} \frac{V}{\delta} \nonumber\\
&+& \mathcal{O}\left((V/\delta)^3\right), \\
\Delta P_{e\tau}^{\rm CPT} &\simeq& -8 s_{12}^2 c_{12}^2 s_{23}^2 \cos
2\theta_{12} \left( \delta L \cos\frac{\delta L}{2} - 2 \sin\frac{\delta L}{2}
\right) \sin\frac{\delta L}{2} \frac{V}{\delta} \nonumber\\
&+& 16 s_{12} c_{12}^3 s_{13} s_{23} c_{23} \cos \delta_{\rm CP} \cos
2\theta_{12} \left( \delta L \cos\frac{\delta L}{2} - 2 \sin\frac{\delta
  L}{2} \right) \sin\frac{\delta L}{2} \frac{V}{\delta} \nonumber\\
&+& 16 s_{12} c_{12} s_{13} s_{23} c_{23} \sin \delta_{\rm CP} 
\Bigg\{ \cos 2\theta_{12} \bigg[ \delta L \cos \delta L - \cos \Delta
  L \nonumber\\
&\times& \left( \delta L \cos\frac{\delta L}{2} - 2 \sin\frac{\delta L}{2}
\right) - \sin \delta L \bigg] + \delta L \sin\frac{\delta L}{2} \sin
\Delta L \Bigg\} \frac{V}{\delta} \nonumber\\
&+& \mathcal{O}\left((V/\delta)^3\right),\\
\Delta P_{\mu e}^{\rm CPT} &\simeq& -8 s_{12}^2 c_{12}^2 c_{23}^2 \cos
2\theta_{12} \left( \delta L \cos\frac{\delta L}{2} - 2 \sin\frac{\delta L}{2}
\right) \sin\frac{\delta L}{2} \frac{V}{\delta} \nonumber\\
&-& 16 s_{12} c_{12}^3 s_{13} s_{23} c_{23} \cos \delta_{\rm CP} \cos
2\theta_{12} \left( \delta L \cos\frac{\delta L}{2} - 2 \sin\frac{\delta
  L}{2} \right) \sin\frac{\delta L}{2} \frac{V}{\delta} \nonumber\\
&+& 16 s_{12} c_{12} s_{13} s_{23} c_{23} \sin \delta_{\rm CP} 
\Bigg\{ \cos 2\theta_{12} \bigg[ \delta L \cos \delta L - \cos \Delta
  L \nonumber\\
&\times& \left( \delta L \cos\frac{\delta L}{2} - 2 \sin\frac{\delta L}{2}
\right) - \sin \delta L \bigg] + \delta L \sin\frac{\delta L}{2} \sin
\Delta L \Bigg\} \frac{V}{\delta} \nonumber\\
&+& \mathcal{O}\left((V/\delta)^3\right), \\
\Delta P_{\tau e}^{\rm CPT} &\simeq& -8 s_{12}^2 c_{12}^2 s_{23}^2 \cos
2\theta_{12} \left( \delta L \cos\frac{\delta L}{2} - 2 \sin\frac{\delta L}{2}
\right) \sin\frac{\delta L}{2} \frac{V}{\delta} \nonumber\\
&+& 16 s_{12} c_{12}^3 s_{13} s_{23} c_{23} \cos \delta_{\rm CP} \cos
2\theta_{12} \left( \delta L \cos\frac{\delta L}{2} - 2 \sin\frac{\delta
  L}{2} \right) \sin\frac{\delta L}{2} \frac{V}{\delta} \nonumber\\
&-& 16 s_{12} c_{12} s_{13} s_{23} c_{23} \sin \delta_{\rm CP} 
\Bigg\{ \cos 2\theta_{12} \bigg[ \delta L \cos \delta L - \cos \Delta
  L \nonumber\\
&\times& \left( \delta L \cos\frac{\delta L}{2} - 2 \sin\frac{\delta L}{2}
\right) - \sin \delta L \bigg] + \delta L \sin\frac{\delta L}{2} \sin
\Delta L \Bigg\} \frac{V}{\delta} \nonumber\\
&+& \mathcal{O}\left((V/\delta)^3\right). \label{eq:DPteCPTlow}
\end{eqnarray}
Note that there are, of course, no terms in the CPT probability
differences that are constant in the matter potential $V$, since in
the limit $V \to 0$, \ie, in vacuum, the CPT probability differences
must vanish, because in vacuum they are equal to zero [\cf,
\eq~(\ref{eq:CPTvacuum})]. Furthermore, we observe that the
leading order terms in the CPT probability differences are linear in the
matter potential $V$, whereas the next-to-leading order terms are
cubic, \ie, there are no second order terms. However, we do not show
the explicit forms of the cubic terms, since they are quite lengthy.
Actually, for symmetric matter density profiles it holds that
the oscillation transition probabilities in matter for
neutrinos and antineutrinos, $P(\nu_\alpha \to \nu_\beta; V)$ and
$P(\bar\nu_\alpha \to \bar\nu_\beta; V)$, respectively, are related by
$P(\nu_\alpha \to \nu_\beta; V) = P(\bar\nu_\beta \to \bar\nu_\alpha; -V)$ [see
\Ref~\cite{Minakata:2002qe} and \eqs~(\ref{eq:prob_nu}) and
(\ref{eq:prob_antinu})]. Hence, in this 
case, the CPT probability differences $\Delta P_{\alpha\beta}^{\rm
CPT}(V) = P(\nu_\alpha \to \nu_\beta; V) - P(\bar\nu_\beta \to
\bar\nu_\alpha; V) = P(\nu_\alpha \to \nu_\beta; V) - P(\nu_\alpha
\to \nu_\beta; -V) \equiv f(V) - f(-V)$ are always odd functions with
respect to the (symmetric) matter potential $V$, since $\Delta
P_{\alpha\beta}^{\rm CPT}(-V) = f(-V) - f(V) = - [f(V) - f(-V)] = -
\Delta P_{\alpha\beta}^{\rm CPT}(V)$ \cite{Minakata:2003}.

Introducing the Jarlskog invariant \cite{Jarlskog:1985ht,Jarlskog:1985cw}
\begin{equation}
J \equiv s_{12} c_{12} s_{13} c_{13}^2 s_{23} c_{23} \sin \delta_{\rm CP}
\simeq s_{12} c_{12} s_{13} s_{23} c_{23} \sin \delta_{\rm CP},
\end{equation}
we can, \eg, write the CPT probability difference $\Delta P_{e\mu}^{\rm
  CPT}$ as
\begin{eqnarray}
\Delta P_{e\mu}^{\rm CPT} &\simeq& - c_{23}^2 \Delta P_{ee}^{\rm CPT}
\nonumber\\
&-& 16 c_{12}^2 \cos 2\theta_{12} J \cot \delta_{\rm CP} \left( \delta
L \cos \frac{\delta L}{2} - 2 \sin \frac{\delta L}{2} \right) \sin
\frac{\delta L}{2} \frac{V}{\delta} \nonumber\\
&-& 16 J \Bigg\{ \cos 2\theta_{12} \bigg[ \delta L \cos \delta L -
  \cos \Delta L \left( \delta L \cos \frac{\delta L}{2} - 2 \sin
  \frac{\delta L}{2} \right) \nonumber\\
&-& \sin \delta L \bigg] + \delta L \sin \frac{\delta L}{2} \sin
\Delta L \Bigg\} \frac{V}{\delta} + \mathcal{O}\left((V/\delta)^3\right).
\end{eqnarray}
In the case of maximal solar mixing, \ie, if the solar mixing angle
$\theta_{12} = \tfrac{\pi}{4}$, then we have
\begin{equation}
\Delta P_{ee}^{\rm CPT} \simeq 0,
\end{equation}
which is also obtained using \eq~(\ref{eq:Pee}), and
\begin{equation}
\Delta P_{e\mu}^{\rm CPT} \simeq - 16 J \delta L \sin \frac{\delta
  L}{2} \sin \Delta L \frac{V}{\delta} \simeq - \Delta P_{\mu e}^{\rm CPT},
\end{equation}
where in this case $J = \tfrac{1}{2} s_{13} s_{23} c_{23} \sin
\delta_{\rm CP}$.
Thus, we would not be able to observe any extrinsic CPT violation in
the $\nu_e \to \nu_e$ and $\bar\nu_e \to \bar\nu_e$ channels. However,
it would still be possible to do so in the $\nu_e \to \nu_\mu$ and
$\bar\nu_e \to \bar\nu_\mu$ channels.
Furthermore, note that if in addition $\delta_{\rm CP} = 0$, then
also $\Delta P_{e\mu}^{\rm CPT}$ and $\Delta P_{\mu e}^{\rm CPT}$
vanish, since $J \propto \sin \delta_{\rm CP}$.

\subsubsection{Step-function matter density profiles}

Next, we consider step-function matter density profiles, \ie, matter
density profiles consisting of two different layers of constant
densities. Let the widths of the two layers be $L_1$ and $L_2$,
respectively, and the corresponding matter potential $V_1$ and $V_2$.
Furthermore, we again let $E_\nu$ denote the neutrino energy.
Similar to the constant matter density profile case, we define the quantities
\begin{eqnarray}
\omega_i &\equiv& \frac{\delta}{2} \sqrt{\left[ \cos 2\theta_{12} -
\frac{V_i}{\delta} \right]^2 + \sin^2 2\theta_{12}},\\
\bar\omega_i &\equiv& \frac{\delta}{2} \sqrt{\left[ \cos 2\theta_{12} +
\frac{V_i}{\delta} \right]^2 + \sin^2 2\theta_{12}},\\
\tilde\Delta_i &=& \Delta - \frac{1}{2} (V_i + \delta) = \frac{\delta}{2}
\left( 2 \frac{\Delta}{\delta} - 1 - \frac{V_i}{\delta} \right),\\
\bar{\tilde\Delta}_i &=& \Delta - \frac{1}{2} (- V_i + \delta) =
\frac{\delta}{2}
\left( 2 \frac{\Delta}{\delta} - 1 + \frac{V_i}{\delta} \right),\\
\theta_{m,i} &\equiv& \frac{1}{2} \arccos \left( \frac{\delta \cos
  2\theta_{12} - V_i}{2\omega_i} \right),\\
\bar\theta_{m,i} &\equiv& \frac{1}{2} \arccos \left( \frac{\delta \cos
  2\theta_{12} + V_i}{2\bar\omega_i} \right),
\end{eqnarray}
with $i = 1,2$ denoting the two different layers,
where again $\delta \equiv \frac{\Delta m_{21}^2}{2 E_\nu}$, $\Delta \equiv
\frac{\Delta m_{31}^2}{2 E_\nu}$, and $\theta_{12}$ is the solar
mixing angle. We divide the time
interval of the neutrino evolution into two parts: $0 \leq t < L_1$
and $L_1 \leq t \leq L$, where $L \equiv L_1 + L_2$. In the first
interval, the parameters $\alpha$, $\bar\alpha$, $\beta$, $\bar\beta$,
$f$, and $\bar f$ are given by the well-known evolution in constant
matter density, \ie, by \eqs~(\ref{eq:alphat0}) - (\ref{eq:barft0})
with the replacements $\omega \to \omega_1$, $\bar\omega \to
\bar\omega_1$, $\tilde\Delta \to \tilde\Delta_1$, $\bar{\tilde\Delta}
\to \bar{\tilde\Delta}_1$, $\theta_m \to \theta_{m,1}$, and $\bar\theta_m \to
\bar\theta_{m,1}$, whereas in the second interval, they are given by
\begin{eqnarray}
\alpha(t,t_0) &=& c_1 c_2' - s_1 s_2' \cos (2\theta_{m,1} -
2\theta_{m,2}) \nonumber\\
&+& {\rm i} (s_1 c_2' \cos 2\theta_{m,1} + s_2' c_1 \cos
2\theta_{m,2}), \\
\bar\alpha(t,t_0) &=& \bar c_1 \bar c_2' - \bar s_1 \bar s_2' \cos
(2\bar\theta_{m,1} - 2\bar\theta_{m,2}) \nonumber\\
&+& {\rm i} (\bar s_1 \bar c_2' \cos 2\bar\theta_{m,1} + \bar s_2'
\bar c_1 \cos 2\bar\theta_{m,2}), \\
\beta(t,t_0) &=& s_1 s_2' \sin (2\theta_{m,1} - 2\theta_{m,2}) - {\rm
  i} (s_1 c_2' \sin 2\theta_{m,1} + s_2' c_1 \sin 2\theta_{m,2}), \\
\bar\beta(t,t_0) &=& \bar s_1 \bar s_2' \sin (2\bar\theta_{m,1} -
2\bar\theta_{m,2}) - {\rm i} (\bar s_1 \bar c_2' \sin 2\bar\theta_{m,1}
+ \bar s_2' \bar c_1 \sin 2\bar\theta_{m,2}), \\
f(t,t_0) &=& {\rm e}^{-{\rm i} \left[ \tilde\Delta_1 (L_1 - t_0) +
  \tilde\Delta_2 (t - L_1) \right]}, \\
\bar f(t,t_0) &=& {\rm e}^{-{\rm i} \left[ \bar{\tilde\Delta}_1 (L_1 - t_0) +
  \bar{\tilde\Delta}_2 (t - L_1) \right]},
\end{eqnarray} 
where $s_i \equiv \sin \omega_i L_i$, $c_i
\equiv \cos \omega_i L_i$, $\bar s_i \equiv \sin \bar\omega_i L_i$,
and $\bar c_i = \cos \bar\omega_i L_i$ with $i = 1,2$ and $s_2' \equiv
\sin \omega_2 \tau$, $c_2' \equiv \cos \omega_2 \tau$, $\bar 
s_2' \equiv \sin \bar\omega_2 \tau$, and $\bar c_2' \equiv \cos \bar\omega_2
\tau$ with $\tau = t - L_1$.

Now, we will take a quick look at the general way of deriving
expressions for the CPT probability differences for the step-function
matter density profile. However, in this case, the derivations are
quite cumbersome and we will only present the results for the CPT
probability difference $\Delta P_{ee}^{\rm CPT}$.

Similar to the case of constant matter density, we obtain the CPT
probability difference $\Delta P_{ee}^{\rm CPT}$ for step-function
matter density profiles as
\begin{eqnarray}
\Delta P_{ee}^{\rm CPT} &\simeq& (s_1 c_2 \cos 2\theta_{m,1} + s_2 c_1
\cos 2\theta_{m,2})^2 + \left[c_1 c_2 - s_1 s_2 \cos 2(\theta_{m,1} -
\theta_{m,2})\right]^2 \nonumber\\
&-& (\bar s_1 \bar c_2 \cos 2\bar\theta_{m,1} +
\bar s_2 \bar c_1 \cos 2\bar\theta_{m,2})^2 + \left[\bar c_1 \bar c_2
  - \bar s_1 \bar s_2 \cos 2(\bar\theta_{m,1} -
  \bar\theta_{m,2})\right]^2. \nonumber\\
\end{eqnarray}
In the low-energy region $V_{1,2} \lesssim \delta \ll \Delta$, we find that 
\begin{eqnarray}
\Delta P_{ee}^{\rm CPT} &\simeq& 8 s_{12}^2 c_{12}^2 \cos 2\theta_{12}
\bigg[ \delta \left( L_1 \frac{V_1}{\delta} + L_2 \frac{V_2}{\delta} \right)
  \cos \frac{\delta (L_1 + L_2)}{2} \nonumber\\
&-& 2 \left( \frac{V_1}{\delta} \sin \frac{\delta L_1}{2} \cos
  \frac{\delta L_2}{2} + \frac{V_2}{\delta} \sin \frac{\delta L_2}{2}
  \cos \frac{\delta L_1}{2} \right) \bigg] \sin \frac{\delta (L_1 +
  L_2)}{2} \nonumber\\
&+& \mathcal{O}\left( (V_1/\delta)^2, (V_2/\delta)^2, V_1 V_2/\delta^2
\right).
\end{eqnarray}
One observes that the CPT probability difference $\Delta P_{ee}^{\rm
CPT}$ is completely symmetric with respect to the exchange of layers 1
and 2. Furthermore, in the limit $V_{1,2} \to V$ and $L_{1,2} \to
L/2$, one recovers the CPT probability difference for constant matter
density (as one should), see \eq~(\ref{eq:DPeeCPTlow}).

\section{Numerical Analysis and Implications for Neutrino
  Oscillation Experiments}
\label{sec:numcal}

In general, the three flavor neutrino oscillation transition probabilities in
matter $P_{\alpha\beta} \equiv P(\nu_\alpha \to \nu_\beta)$ are
complicated (mostly trigonometric) functions depending on nine parameters
\begin{equation}
P_{\alpha\beta} = P_{\alpha\beta}(\Delta m_{21}^2, \Delta m_{31}^2,
\theta_{12}, \theta_{13}, \theta_{23}, \delta_{\rm CP}; E_\nu, L,
V(L)), \quad \alpha,\beta = e,\mu,\tau,
\end{equation}
where $\Delta m_{21}^2$ and $\Delta m_{31}^2$ are the neutrino mass squared
differences, $\theta_{12}$, $\theta_{13}$, $\theta_{23}$, and
$\delta_{\rm CP}$ are the leptonic mixing parameters, $E_\nu$ is the
neutrino energy, $L$ is the baseline length, and finally, $V(L)$ is the matter
potential, which generally depends on $L$. Naturally,
the CPT probability differences depend on the same parameters as the
neutrino oscillation transition probabilities.
The neutrino mass squared differences and the leptonic mixing
parameters are fundamental parameters given by Nature, and thus, do
not vary in any experimental setup, whereas the neutrino energy,
the baseline length, and the matter potential depend on the specific
experiment that is studied.

The present values of the fundamental neutrino parameters are given in
Table~\ref{tab:parameters}. 
\begin{table}[t!]
\begin{center}
\caption{\label{tab:parameters} Present values of the fundamental
  neutrino parameters.}
\begin{tabular}{cccc}
\hline
Parameter & Best-fit value & Range & References\\
\hline
$\Delta m_{21}^2$ & $7.1 \cdot 10^{-5} \, {\rm eV}^2$ & {\scriptsize
  $\sim (6 \div 9) \cdot 10^{-5} \, {\rm eV}^2$ (99.73~\% C.L.)} &
\cite{Fukuda:2001nj,Fukuda:2001nk,Smy:2001wf,Fukuda:2002pe,Smy:2002,Ahmad:2001an,Ahmad:2002jz,Ahmad:2002ka,Hallin:2002,Bahcall:2001zu,Bahcall:2001cb,Bahcall:2002hv,Bahcall:2002ij}\\
$|\Delta m_{31}^2|$ & $2.5 \cdot 10^{-3} \, {\rm eV}^2$ & {\scriptsize
  $(1.6 \div 3.9) \cdot 10^{-3} \, {\rm eV}^2$ (90~\% C.L.)} &
\cite{Fukuda:1998mi,Fukuda:1998ah,Fukuda:2000np,Shiozawa:2002}\\
$\theta_{12}$ & $34^\circ$ & $27^\circ \div 44^\circ$ (99.73~\% C.L.) &
\cite{Fukuda:2001nj,Fukuda:2001nk,Smy:2001wf,Fukuda:2002pe,Smy:2002,Ahmad:2001an,Ahmad:2002jz,Ahmad:2002ka,Hallin:2002,Bahcall:2001zu,Bahcall:2001cb,Bahcall:2002hv,Bahcall:2002ij}\\
$\theta_{13}$ & - & $0 \div 9.2^\circ$ (90~\% C.L.) &
\cite{Apollonio:1998xe,Apollonio:1999ae,Apollonio:2002gd}\\
$\theta_{23}$ & $45^\circ$ & $37^\circ \div 45^\circ$ (90~\% C.L.) &
\cite{Fukuda:1998mi,Fukuda:1998ah,Fukuda:2000np,Shiozawa:2002}\\
$\delta_{\rm CP}$ & - & $[0,2\pi)$ & - \\
\hline
\end{tabular}
\end{center}
\end{table}
These values are motivated by recent
global fits to different kinds of neutrino oscillation data. All
results within this study are, unless otherwise stated, calculated for
the best-fit values given in Table~\ref{tab:parameters}. Furthermore,
we assume a normal neutrino mass hierarchy spectrum, \ie, $\Delta
m_{21}^2 \ll \Delta m_{31}^2$ with $\Delta m_{31}^2 = +
2.5 \cdot 10^{-3} \, {\rm eV}^2$. For the leptonic mixing angle
$\theta_{13}$, we only allow values below the CHOOZ upper bound, \ie,
$\sin^2 2\theta_{13} \lesssim 0.1$ or $\theta_{13} \lesssim
9.2^\circ$. For the CP violation phase, we use different values
between $0$ and $2\pi$, \ie, $\delta_{\rm CP} \in [0,2\pi)$. Note that
there is no CP violation if $\delta_{\rm CP} \in \{0,\pi\}$, whereas
the effects of CP violation are maximal if $\delta_{\rm CP} \in
\{\tfrac{\pi}{2},\tfrac{3\pi}{2}\}$. 

As realistic examples, let us now investigate the effects of extrinsic CPT
violation on the transition probabilities for neutrino oscillations in
matter for various experiment in which the neutrinos traverse the
Earth. Such experiments are \eg~so-called long baseline experiments,
atmospheric and solar neutrino oscillation experiments. In some
analyses of these experiments, the Preliminary Reference Earth Model
(PREM) matter density profile \cite{Dziewonski:1981xy} has been used,
which has been obtained from geophysics using seismic wave
measurements. However, the (mantle-core-mantle) step-function matter
density profile\footnote{The 
  mantle-core-mantle step-function approximation of the Earth matter
  density profile consists of three constant matter density layers
  (mantle, core, and mantle) with $\rho_{\rm mantle} = 4.5 \, {\rm
  g/cm^3}$ and $\rho_{\rm core} = 11.5 \, {\rm g/cm^3}$.} is an
excellent approximation to the PREM matter density profile
\cite{Freund:1999vc}, whereas the constant matter density profile
serves as a very good approximation to long baseline experiments that
have baselines that do not enter the core of the Earth. Thus, we use
these approximations for our calculations.

The equatorial radius of the Earth and the radius of the core of the
Earth are $R_\oplus \simeq 6371 \, {\rm km}$ and $r \simeq 3486 \,
{\rm km}$, respectively, which means that the thickness of the mantle
of the Earth is $R_\oplus - r \simeq 2885 \, {\rm km}$. From the
geometry of the Earth, one finds that the relation between the maximal
depth of the baseline $\ell$ and the baseline length $L$ is given by
\begin{equation}
\ell = R_\oplus - \sqrt{R_\oplus^2 - \frac{L^2}{4}} \quad \mbox{(or $L
= 2 \sqrt{\ell(2R_\oplus - \ell)}$)}.
\end{equation}
Hence, in order for the neutrinos also to traverse the core of the
Earth, \ie, $\ell \geq R_\oplus - r$, the baseline length needs to be
$L \gtrsim 10670 \, {\rm km}$. This means that for experiments with
baseline lengths shorter than $10670 \, {\rm km}$, we can safely use
the constant matter density profile. For ``shorter'' long baseline
experiments ($L \lesssim 3000 \, {\rm km}$) we use the average matter
density of the continental Earth crust, $\rho_{\rm crust} \simeq 3 \,
{\rm g/cm^3}$, whereas for ``longer'' long baseline experiments ($3000
\, {\rm km} \lesssim L \lesssim 10670 \, {\rm km}$) we use the average
matter density of the mantle of the Earth, $\rho_{\rm mantle} \simeq
4.5 \, {\rm g/cm^3}$.
Furthermore, the matter potential $V \equiv V(L)$ expressed in terms
of the matter density $\rho \equiv \rho(L)$ is given by
\begin{equation}
V \simeq \frac{1}{\sqrt{2}} G_F \frac{1}{m_N} \rho \simeq 3.78
\cdot 10^{-14} \, {\rm eV} \cdot \rho[{\rm g/cm^3}],
\label{eq:Vformel}
\end{equation}
where $\rho[{\rm g/cm^3}]$ is the matter density given in units of
${\rm g/cm^3}$.

Let us now investigate when it is possible to use the low-energy
approximations for the CPT probability differences derived in the
previous section. In these approximations, we have assumed that the
matter potential $V$ is smaller than the parameter $\delta$, \ie, $V
\lesssim \delta \ll \Delta$. Now, the parameter $\delta$ is a function
of the neutrino energy $E_\nu$:
\begin{equation}
\delta = \frac{\Delta m_{21}^2}{2E_\nu} \simeq 3.55 \cdot 10^{-5} \,
{\rm eV} \cdot E_\nu[{\rm eV}]^{-1},
\label{eq:dformel}
\end{equation}
where $E_\nu[{\rm eV}]$ is the neutrino energy in eV. Thus, combining
\eqs~(\ref{eq:Vformel}) and (\ref{eq:dformel}), we find that
\begin{equation}
E_\nu \lesssim 0.94 \cdot 10^9 \, {\rm eV} \cdot \rho[{\rm g/cm^3}]^{-1},
\end{equation}
which means that for the continental Earth crust ($\rho_{\rm crust}
\simeq 3 \, {\rm g/cm^3}$) the neutrino energy $E_\nu$ must be smaller
than about $0.31 \, {\rm GeV}$ in order for the low-energy
approximations to be valid.

In Table~\ref{tab:lbl}, we list several past, present, and future long
baseline experiments of accelerator and reactor types including their
specific parameter sets for which we are going to estimate the extrinsic
CPT violation effects.
\begin{table}[ph!]
\begin{center}
\caption{\label{tab:lbl} Accelerator and reactor long baseline
  experiments including measurable neutrino oscillation channels, average
  neutrino energies ($E_\nu$), approximate baseline lengths ($L$) as well as
  references to the respective experiments. The CHOOZ, KamLAND, and
  Palo Verde experiments are reactor experiments, whereas the other
  experiments are accelerator experiments. Furthermore, the BooNE,
  MiniBooNE, CHOOZ, LSND, NuTeV, and Palo Verde experiments are
  sometimes called short baseline experiments. However, we will use
  the term long baseline experiments for all experiments in this
  table.}
\begin{tabular}{lcccc}
\hline
Experiment & Channels & $E_\nu$ & $L$ & References\\
\hline
BNL NWG & $\nu_\mu \to \nu_e$ & $1 \, {\rm GeV}$ & $400
\, {\rm km}$, $2540 \, {\rm km}$ &
\cite{Beavis:2002ye,Marciano:2001tz,Diwan:2003bp}\\
BooNE & $\left\{ \begin{array}{c} \nu_\mu \to \nu_e\\
\bar\nu_\mu \to \bar\nu_e \end{array} \right.$
& $(0.5 \div 1.5) \, {\rm GeV}$ & $1 \, {\rm km}$ & \cite{Zimmerman:2002xj} \\
MiniBooNE & $\left\{ \begin{array}{c} \nu_\mu \to \nu_e\\ \bar\nu_\mu \to
\bar\nu_e \end{array} \right.$ & $(0.5 \div 1.5) \, {\rm GeV}$ & $500
\, {\rm m}$ & \cite{Bazarko:2002pn}\\
CHOOZ & $\bar\nu_e \to \bar\nu_e$ & $\sim 3 \, {\rm MeV}$ &
$1030 \, {\rm m}$ & \cite{Apollonio:1998xe,Apollonio:1999ae,Apollonio:2002gd}\\
ICARUS & $\left\{ \begin{array}{c} \nu_\mu \to \nu_e\\ \nu_\mu \to
  \nu_\tau \end{array} \right.$ & $17 \, {\rm GeV}$ & $743 \, {\rm km}$ &
\cite{Rubbia:1996ip,Arneodo:2001tx,Duchesneau:2002yq}\\
JHF-Kamioka & $\left\{ \begin{array}{c} \nu_\mu \to \nu_e\\ \nu_\mu
\to \nu_\mu \end{array} \right.$
& $(0.4 \div 1.0) \, {\rm GeV}$ & $295 \, {\rm km}$ & \cite{Itow:2001ee}\\
K2K & $\left\{ \begin{array}{c} \nu_\mu \to \nu_e\\ \nu_\mu \to
\nu_\mu \end{array} \right.$ & $1.3 \, {\rm GeV}$ & $250 \, {\rm km}$
& \cite{Ahn:2001cq,Ahn:2002up}\\
KamLAND & $\bar\nu_e \to \bar\nu_e$ & $\sim 3 \, {\rm MeV}$
& $\sim 180 \, {\rm km}$ & \cite{Eguchi:2002dm}\\
LSND & $\left\{ \begin{array}{c} \nu_\mu \to \nu_e\\ \bar\nu_\mu \to
\bar\nu_e \end{array} \right.$ & $48 \, {\rm MeV} $ & $30 \, {\rm m}$ &
\cite{Athanassopoulos:1996jb,Athanassopoulos:1998pv,Aguilar:2001ty}\\
MINOS & $\left\{ \begin{array}{c} \nu_\mu \to \nu_e\\ \nu_\mu \to
\nu_\mu \end{array} \right.$ & $(3 \div 18) \, {\rm GeV}$ &
$735 \, {\rm km}$ & \cite{Wojcicki:2001ts,Paolone:2001am,Diwan:2002pu}\\
NuMI I/II & $\left\{ \begin{array}{c} \nu_\mu \to \nu_e\\ \bar\nu_\mu
\to \bar\nu_e \end{array} \right.$ & $1.4 \, {\rm GeV}$ / $0.7 \, {\rm
GeV}$ & $712 \, {\rm km}$ / $987 \, {\rm km}$ & \cite{Ayres:2002nm}\\
NuTeV & $\left\{ \begin{array}{c} \nu_\mu \to \nu_e\\ \bar\nu_\mu \to
\bar\nu_e \end{array} \right.$  & $75 \, {\rm GeV}$, $200 \, {\rm
GeV}$ & $(915 \div 1235) \, {\rm m}$ & \cite{Avvakumov:2002jj}\\
OPERA & $\nu_\mu \to \nu_\tau$ & $17 \, {\rm GeV}$ &
$743 \, {\rm km}$ & \cite{Pessard:2001vs,Duchesneau:2002yq}\\
Palo Verde & $\bar\nu_e \to \bar\nu_e$ & $\sim 3 \, {\rm
MeV}$ & $750 \, {\rm m}$, $890 \, {\rm m}$ &
\cite{Boehm:1998zp,Boehm:1999gk,Boehm:2000vp,Boehm:2001ik}\\
\hline
\end{tabular}
\end{center}
\end{table}
\begin{table}[h!]
\begin{center}
\caption{\label{tab:CPTlbl} Estimates of the CPT probability
  differences for the different long baseline experiments listed in
  Table~\ref{tab:lbl}. The fundamental neutrino parameters used are: 
  $\Delta m_{21}^2 = 7.1 \cdot 10^{-5} \, {\rm eV}^2$, $\Delta
  m_{31}^2 = 2.5 \cdot 10^{-3} \, {\rm eV}^2$, $\theta_{12} =
  34^\circ$, $\theta_{13} = 9.2^\circ$, $\theta_{23} = 45^\circ$,
  and $\delta_{\rm CP} = 0$. Furthermore, we have used constant
  matter density profiles with $\rho = 3 \, {\rm g/cm^3}$ as
  approximations of the continental Earth crust.}
\begin{tabular}{lcccc}
\hline
Experiment & \multicolumn{4}{c}{CPT probability differences}\\
 & Quantities & Numerical & Analytical & Analytical (low-energy)\\
\hline
BNL NWG & $\Delta P_{\mu e}^{\rm CPT}$ & $0.010$ & $3.6 \cdot
10^{-4}$ & $1.7 \cdot 10^{-6}$\\
BNL NWG & $\Delta P_{\mu e}^{\rm CPT}$ & $0.032$ & $1.2 \cdot 10^{-3}$
& $2.7 \cdot 10^{-3}$\\
BooNE & $\Delta P_{\mu e}^{\rm CPT}$ & $6.6 \cdot 10^{-13}$ & $5.1
\cdot 10^{-14}$ & $2.0 \cdot 10^{-17}$\\
MiniBooNE & $\Delta P_{\mu e}^{\rm CPT}$ & $4.1 \cdot 10^{-14}$ &
$3.2 \cdot 10^{-15}$ & $-2.0 \cdot 10^{-17}$\\
CHOOZ & $\Delta P_{ee}^{\rm CPT}$ & $-3.6 \cdot 10^{-5}$ & $-3.7 \cdot
10^{-9}$ & $-3.7 \cdot 10^{-9}$\\
ICARUS & $\Delta P_{\mu e}^{\rm CPT}$ & $4.0 \cdot 10^{-5}$ & $3.1
\cdot 10^{-6}$ & $4.1 \cdot 10^{-9}$\\
 & $\Delta P_{\mu\tau}^{\rm CPT}$ & $-3.8 \cdot 10^{-5}$ & - & -\\
JHF-Kamioka & $\Delta P_{\mu e}^{\rm CPT}$ & $3.8 \cdot 10^{-3}$ &
$2.2 \cdot 10^{-4}$ & $5.0 \cdot 10^{-7}$\\
 & $\Delta P_{\mu\mu}^{\rm CPT}$ & $-1.3 \cdot 10^{-4}$ & - & -\\
K2K & $\Delta P_{\mu e}^{\rm CPT}$ & $1.0 \cdot 10^{-3}$ & $7.2 \cdot
10^{-5}$ & $1.2 \cdot 10^{-7}$\\
 & $\Delta P_{\mu\mu}^{\rm CPT}$ & $-5.3 \cdot 10^{-5}$ & - & -\\
KamLAND & $\Delta P_{ee}^{\rm CPT}$ & $-0.033$ & $-0.040$ & $-0.040$\\
LSND & $\Delta P_{\mu e}^{\rm CPT}$ & $4.8 \cdot 10^{-15}$ & $3.7
\cdot 10^{-16}$ & $1.9 \cdot 10^{-18}$\\
MINOS & $\Delta P_{\mu e}^{\rm CPT}$ & $1.9 \cdot 10^{-4}$ & $1.4
\cdot 10^{-5}$ & $1.9 \cdot 10^{-8}$\\
 & $\Delta P_{\mu\mu}^{\rm CPT}$ & $-1.1 \cdot 10^{-5}$ & - & -\\
NuMI I & $\Delta P_{\mu e}^{\rm CPT}$ & $0.026$ & $-2.7 \cdot 10^{-5}$
& $6.2 \cdot 10^{-6}$\\
NuMI II & $\Delta P_{\mu e}^{\rm CPT}$ & $2.6 \cdot
10^{-3}$ & $-2.4 \cdot 10^{-4}$ & $1.8 \cdot 10^{-4}$\\
NuTeV & $\Delta P_{\mu e}^{\rm CPT}$ & $1.6 \cdot 10^{-18}$ & $1.2
\cdot 10^{-19}$ & $-2.6 \cdot 10^{-15}$\\
NuTeV & $\Delta P_{\mu e}^{\rm CPT}$ & $8.2 \cdot 10^{-20}$ & $6.4
\cdot 10^{-21}$ & $-1.5 \cdot 10^{-15}$\\
OPERA & $\Delta P_{\mu\tau}^{\rm CPT}$ & $-3.8 \cdot 10^{-5}$ & - & -\\
Palo Verde & $\Delta P_{ee}^{\rm CPT}$ & $-1.2 \cdot 10^{-5}$ & $-1.1
\cdot 10^{-9}$ & $-1.1 \cdot 10^{-9}$\\
Palo Verde & $\Delta P_{ee}^{\rm CPT}$ & $-2.2 \cdot
10^{-5}$ & $-2.1 \cdot 10^{-9}$ & $-2.1 \cdot 10^{-9}$\\
\hline
\end{tabular}
\end{center}
\end{table}
From the values of the neutrino energies given in this table we can
conclude that the low-energy approximations for the CPT probability
differences are applicable for the reactor experiments including the
LSND accelerator experiment, but not for the accelerator experiments
in general.

Using the values of the fundamental neutrino parameters given in
Table~\ref{tab:parameters} as well as the approximate values of the neutrino
energy and baseline length for the different long baseline experiments given in
Table~\ref{tab:lbl}, we obtain estimates of the CPT probability
differences, which are presented in Table~\ref{tab:CPTlbl}.
From the values in Table~\ref{tab:CPTlbl} we observe that there are
three different experiments with fairly large estimates of the CPT
probability differences. These experiments are the KamLAND, BNL NWG,
and NuMI experiments, which will later in this paper be studied
in more detail. In general, there is a rather large discrepancy among the
values coming from the numerical, analytical, and low-energy
approximation calculations. This is mainly due to the oscillatory
behavior of the CPT probability differences. Therefore, these values
can change drastically with a small modification of the input
parameter values. Thus, this can explain the somewhat different values
of the different calculations. However, in most of the cases, the
order of magnitude of the different calculations are in
agreement. Note that for all reactor experiments the analytical and
low-energy approximation estimates agree completely, since the
neutrino energies are low enough for these experiments in order for
the low-energy approximations to be valid.
Moreover, we have calculated the CPT probability difference $\Delta
P_{\mu e}^{\rm CPT}$ for two potential neutrino factory setups using
the analytical formula~(\ref{eq:Pme}). In
general, these setups are very long baseline experiments that even
penetrate the Earth's mantle in addition to the Earth's crust. For our
calculations we used a constant matter density profile with $\rho =
\rho_{\rm mantle} \simeq 4.5 \, {\rm g/cm^3}$. Furthermore, we chose
the neutrino energy to be 50~GeV as well as the baseline lengths
3000~km and 7000~km, respectively. For these parameter values, we
obtained $\Delta P_{\mu e}^{\rm CPT} \simeq 3.0 \cdot 10^{-5}$
(3000~km) and $\Delta P_{\mu e}^{\rm CPT} \simeq 1.8 \cdot 10^{-5}$
(7000~km). Thus, the extrinsic CPT violation is practically negligible for a
future neutrino factory.

\begin{figure}[th!]
\begin{center}
\includegraphics[height=14.5cm,angle=-90]{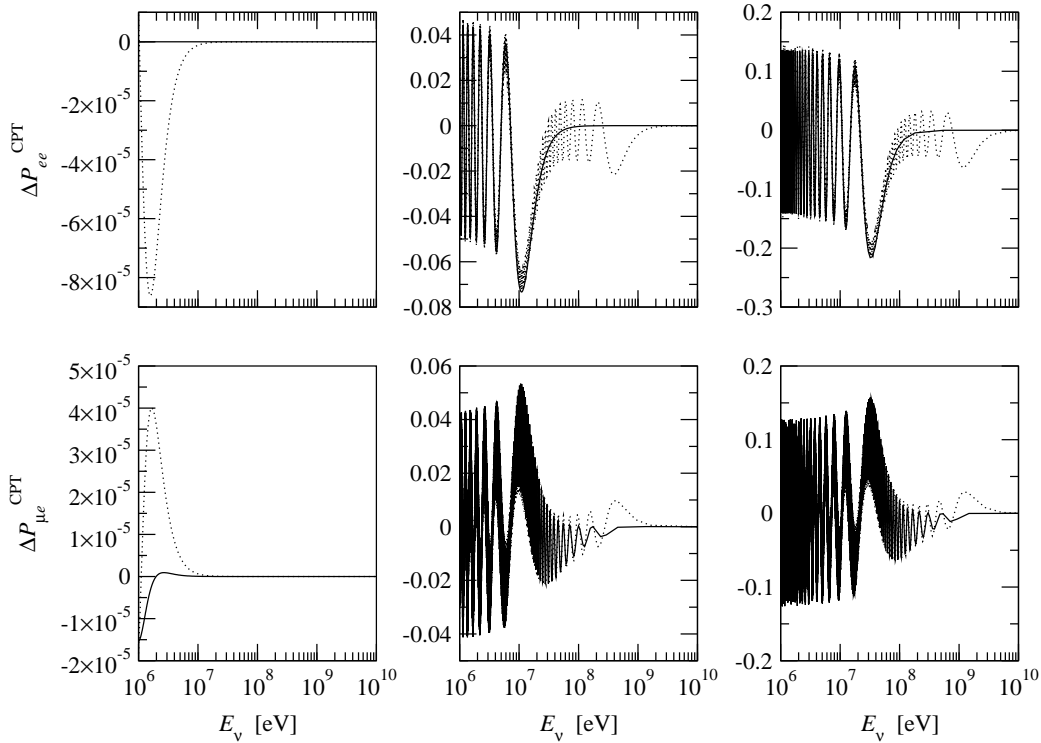}
\end{center}
\caption{\scriptsize The CPT probability differences $\Delta P_{ee}^{\rm
    CPT}$ and $\Delta P_{\mu e}^{\rm CPT}$ plotted as functions of the
    neutrino energy $E_\nu$. The baseline lengths used are: 1~km (left
    column), 250~km (middle column), and 750~km (right column) with
    $\rho = 3 \, {\rm g/cm^3}$. Dotted curves correspond to numerical
    calculations using the evolution operator method and
    Cayley--Hamilton formalism
    \cite{Ohlsson:1999xb,Ohlsson:1999um,Ohlsson:2001vp}, whereas solid
    curves correspond to analytical calculations using
    \eqs~(\ref{eq:Pee}) and (\ref{eq:Pme}). The fundamental neutrino
    parameters used are:  
    $\Delta m_{21}^2 = 7.1 \cdot 10^{-5} \, {\rm eV}^2$, $\Delta
    m_{31}^2 = 2.5 \cdot 10^{-3} \, {\rm eV}^2$, $\theta_{12} =
    34^\circ$, $\theta_{13} = 9.2^\circ$, $\theta_{23} = 45^\circ$,
    and $\delta_{\rm CP} = 0$.}
\label{fig:DeltaPE}
\end{figure}

Next, in \fig~\ref{fig:DeltaPE}, we plot the CPT probability differences
$\Delta P_{ee}^{\rm CPT}$ and $\Delta P_{\mu e}^{\rm CPT}$ as
functions of the neutrino energy $E_\nu$ for three different
characteristic baseline lengths: 1~km, 250~km, and 750~km.
From these plots we observe that the CPT probability differences
increase with increasing baseline length $L$. Furthermore, we note
that for increasing neutrino energy $E_\nu$ the extrinsic CPT
violation effects disappear, since the CPT probability differences go
to zero in the limit when $E_\nu \to \infty$. We also note that
$\Delta P_{ee}^{\rm CPT}$ and $\Delta P_{\mu e}^{\rm CPT}$ are
basically of the same order of magnitude. In this figure, the
numerical curves consist of a modulation of two oscillations: one slow
oscillation with larger amplitude and lower frequency and another fast
oscillation with smaller amplitude and higher frequency. On the other hand, the
analytical curves consist of one oscillation only and they are
therefore not able to reproduce the oscillations with smaller
amplitudes and higher frequencies. However, the agreement between the
two curves are very good considering the oscillations with larger
amplitudes and lower frequencies. In principle, the analytical curves
are running averages of the numerical ones, and in fact, the fast
oscillations cannot be resolved by any realistic detector due to
limited energy resolution making the analytical calculations excellent
approximations of the numerical ones.

\begin{figure}[th!]
\begin{center}
\includegraphics[height=14.5cm,angle=-90]{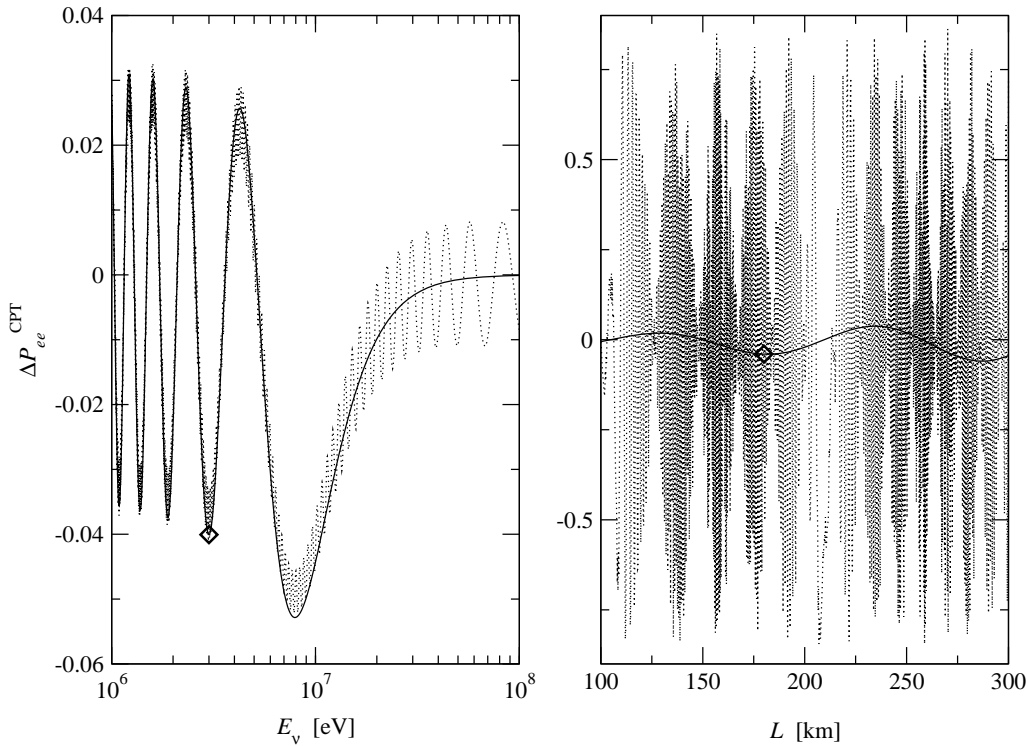}
\end{center}
\caption{\scriptsize The CPT probability difference $\Delta
  P_{ee}^{\rm CPT}$ for 
  the KamLAND experiment. The left-hand side plot shows its dependence on the
  neutrino energy $E_\nu$, whereas the right-hand side plot shows its
  dependence on the baseline length $L$. The solid and dotted curves
  are analytical and numerical results, respectively. The diamonds
  ('$\diamond$') indicate the central values of the KamLAND
  experiment. The parameters used are the same as for \fig~\ref{fig:DeltaPE}.}
\label{fig:KamLAND}
\end{figure}

Let us now investigate some of the most interesting experiments in
more detail for which the extrinsic CPT violation effects may be sizeable.
In \fig~\ref{fig:KamLAND}, we plot the CPT probability
difference $\Delta P_{ee}^{\rm CPT}$ as functions of both the neutrino
energy $E_\nu$ and the baseline length $L$ centered around values of
these parameters characteristic for the KamLAND experiment.
We observe that for neutrino energies around the average neutrino
energy of the KamLAND experiment the CPT probability difference
$\Delta P_{ee}^{\rm CPT}$ could be as large as 3~\% - 5~\% making the
extrinsic CPT violation non-negligible. This means that the transition
probabilities $P(\nu_e \to \nu_e)$ and $P(\bar\nu_e \to \bar\nu_e)$
are not equal to each other for energies and baseline lengths typical
for the KamLAND experiment. Thus, if one would be able to find a
source of electron neutrinos with the same neutrino energy as the
reactor electron antineutrinos coming from the KamLAND experiment,
then one would be able to measure such effects. Furthermore, for the
KamLAND experiment the agreement between the analytical
formula~(\ref{eq:Pee}) and the low-energy
approximation~(\ref{eq:DPeeCPTlow}) is excellent, \ie, it 
is not possible to distinguish the results of these formulas from each
other in the plots. 

\begin{figure}[th!]
\begin{center}
\includegraphics[bb = 230 0 612 792,clip,height=14cm,angle=-90]{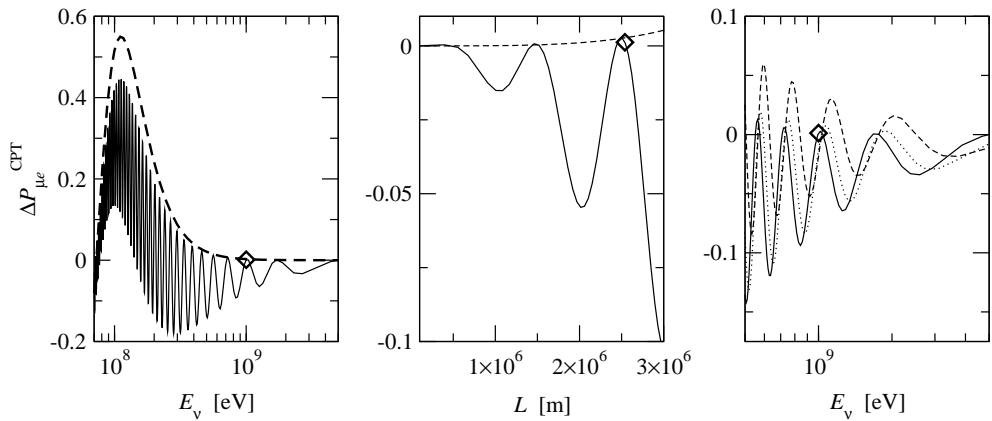}
\end{center}
\caption{\scriptsize The CPT probability difference $\Delta
  P_{\mu e}^{\rm CPT}$ for
  the BNL NWG experiment (baseline length: 2540 km). The left-hand
  side plot shows its dependence on the neutrino energy $E_\nu$ (solid
  curve = analytical calculation, dashed curve = low-energy
  approximation), the middle plot shows its dependence on the baseline
  length $L$ (solid curve = analytical calculation, dashed curve =
  low-energy approximation), and the right-hand side plot shows the
  dependence on $E_\nu$ for three different values of $\delta_{\rm
  CP}$: $0$ (solid curve), $\tfrac{\pi}{4}$ (dotted curve), and
  $\tfrac{\pi}{2}$ (dashed curve). The diamonds ('$\diamond$')
  indicate the central values of the BNL NWG experiment. The other
  parameters used are the same as for \fig~\ref{fig:DeltaPE}.}
\label{fig:BNL}
\end{figure}

\begin{figure}[th!]
\begin{center}
\includegraphics[bb = 230 0 612 792,clip,height=14cm,angle=-90]{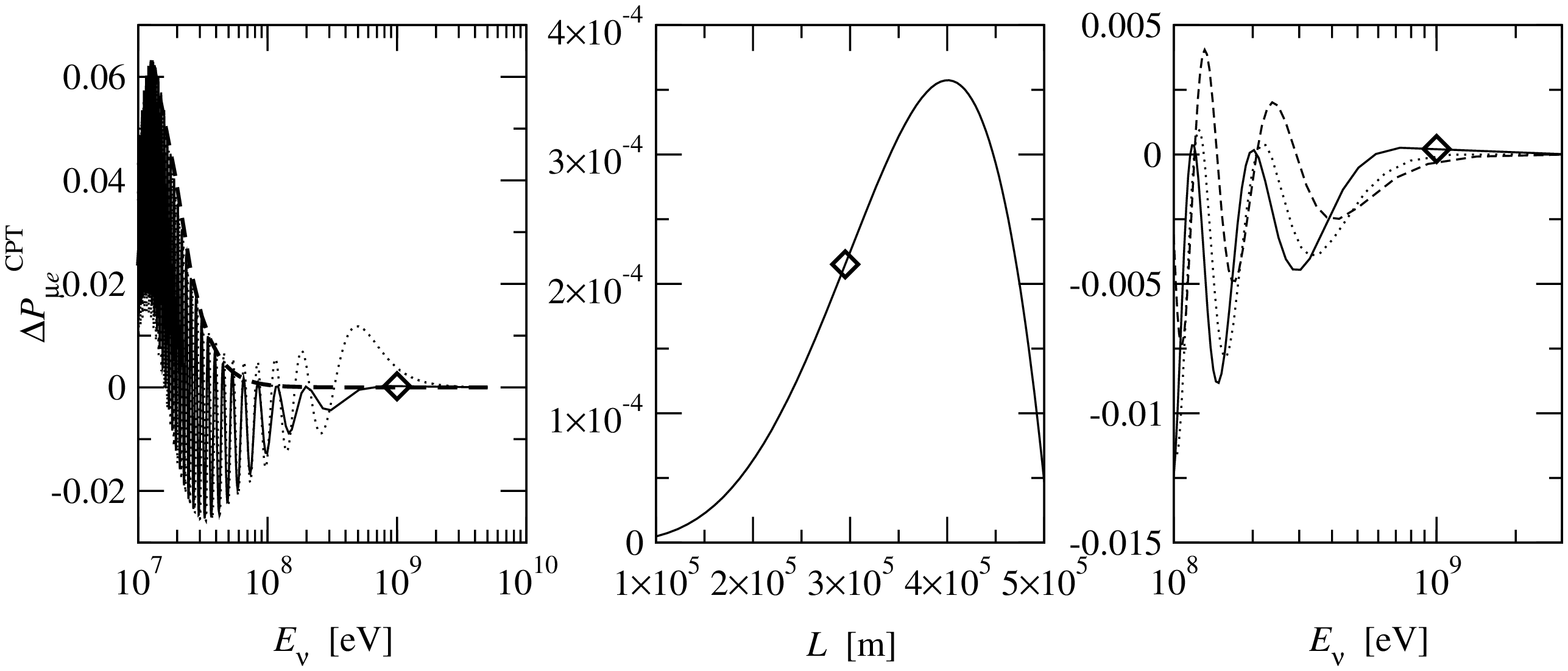}
\end{center}
\caption{\scriptsize The CPT probability difference $\Delta
  P_{\mu e}^{\rm CPT}$ for
  the JHF-Kamioka experiment. The left-hand side plot shows its
  dependence on the neutrino energy $E_\nu$ (dotted curve = numerical
  calculation, solid curve = analytical calculation, and dashed curve
  = low-energy approximation), the middle plot shows its dependence on
  the baseline length $L$ (solid curve = analytical calculation), and
  the right-hand side plot shows the dependence on $E_\nu$ for three
  different values of $\delta_{\rm CP}$: $0$ (solid curve),
  $\tfrac{\pi}{4}$ (dotted curve), and $\tfrac{\pi}{2}$ (dashed
  curve). The diamonds ('$\diamond$') indicate the central values of
  the JHF-Kamioka experiment. The other parameters used are the same
  as for \fig~\ref{fig:DeltaPE}.}
\label{fig:JHF}
\end{figure}

\begin{figure}[th!]
\begin{center}
\includegraphics[bb = 230 0 612 792,clip,height=14cm,angle=-90]{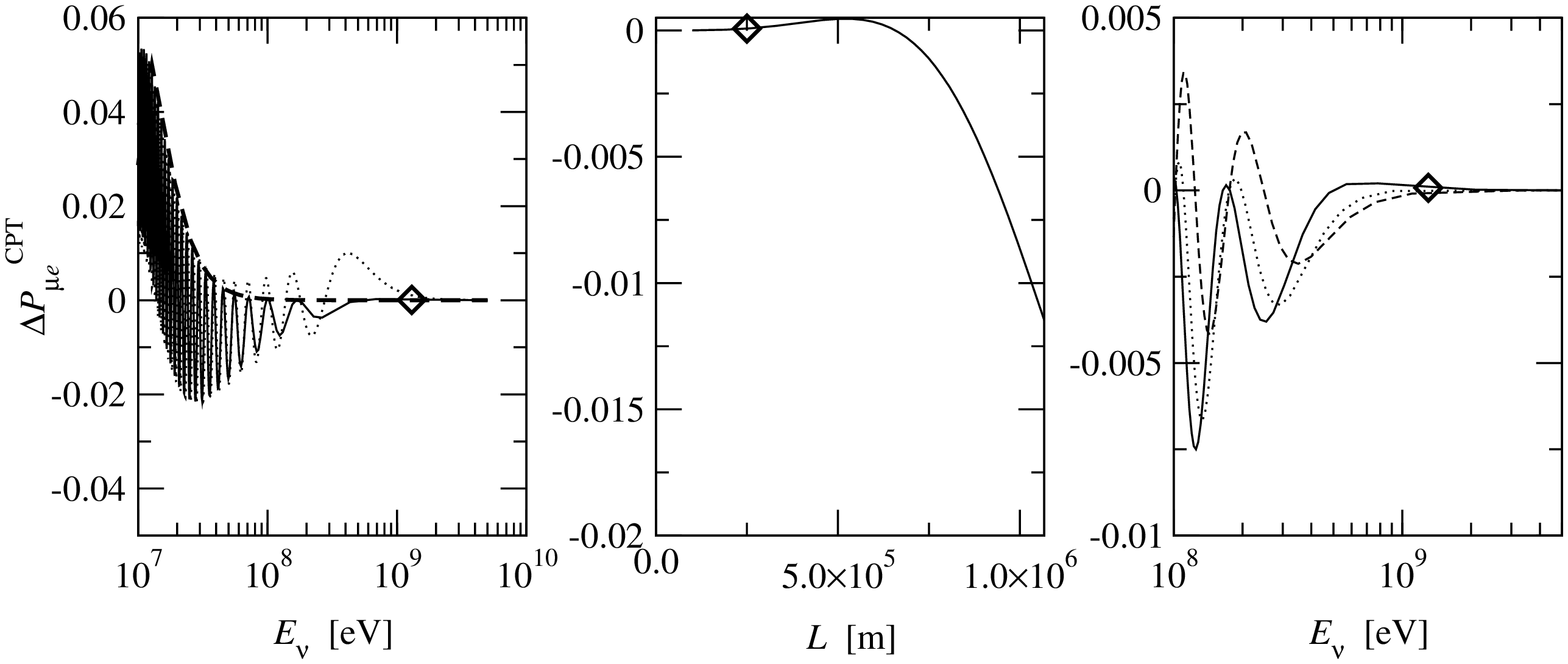}
\end{center}
\caption{\scriptsize The CPT probability difference $\Delta
  P_{\mu e}^{\rm CPT}$ for
  the K2K experiment. The left-hand side plot shows its dependence on
  the neutrino energy $E_\nu$ (dotted curve = numerical calculation,
  solid curve = analytical calculation, and dashed curve = low-energy
  approximation), the middle plot shows its dependence on the baseline
  length $L$ (solid curve = analytical calculation), and the
  right-hand side plot shows the dependence on $E_\nu$ for three
  different values of $\delta_{\rm CP}$: $0$ (solid curve),
  $\tfrac{\pi}{4}$ (dotted curve), and $\tfrac{\pi}{2}$ (dashed
  curve). The diamonds ('$\diamond$') indicate the central values of
  the K2K experiment. The other parameters used are the same as for
  \fig~\ref{fig:DeltaPE}.}
\label{fig:K2K}
\end{figure}

\begin{figure}[th!]
\begin{center}
\includegraphics[bb = 230 0 612 792,clip,height=14cm,angle=-90]{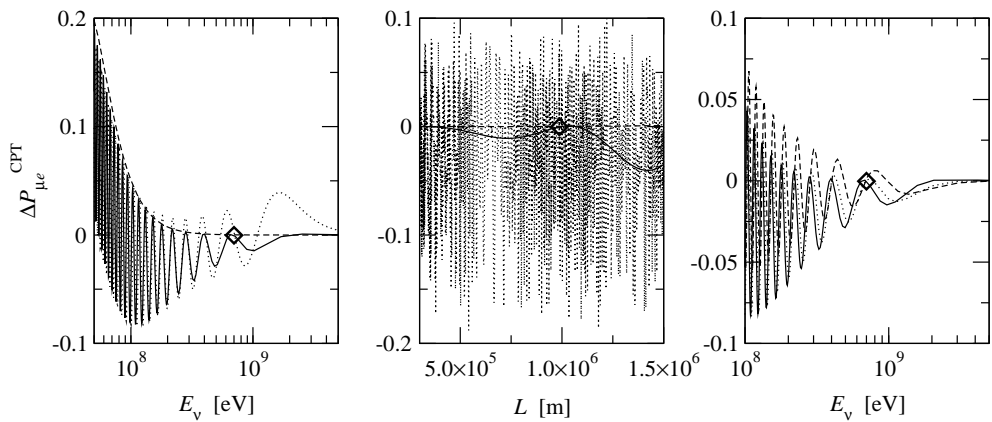}
\end{center}
\caption{\scriptsize The CPT probability difference $\Delta
  P_{\mu e}^{\rm CPT}$ for
  the NuMI Phase II experiment. The left-hand side plot shows its
  dependence on the neutrino energy $E_\nu$ (dotted curve = numerical
  calculation, solid curve = analytical calculation, and dashed curve
  = low-energy approximation), the middle plot shows its dependence on
  the baseline length $L$ (dotted curve = numerical calculation, solid
  curve = analytical calculation, and dashed curve = low-energy
  approximation), and the right-hand side plot shows the dependence on
  $E_\nu$ for three different values of $\delta_{\rm CP}$: $0$ (solid
  curve), $\tfrac{\pi}{4}$ (dotted curve), and $\tfrac{\pi}{2}$
  (dashed curve). The diamonds ('$\diamond$') indicate the central
  values of the NuMI Phase II experiment. The other parameters used
  are the same as for \fig~\ref{fig:DeltaPE}.}
\label{fig:NuMI}
\end{figure}

Next, in \figs~\ref{fig:BNL} - \ref{fig:NuMI}, we present some plots
for the topical accelerator long baseline experiments BNL NWG,
JHF-Kamioka, K2K, and NuMI, which have
approximately the same neutrino energies, but different baseline
lengths. In these figures, we plot the CPT
probability difference $\Delta P_{\mu e}^{\rm CPT}$ as functions of
the neutrino energy $E_\nu$ and the baseline length $L$ as well as the
neutrino energy $E_\nu$ for three different values of the CP violation
phase $\delta_{\rm CP}$ corresponding to no CP violation ($\delta_{\rm
  CP} = 0$), ``intermediate'' CP violation ($\delta_{\rm CP} =
\tfrac{\pi}{4}$), and maximal CP violation ($\delta_{\rm CP} =
\tfrac{\pi}{2}$), respectively. We note that in all cases the
low-energy approximation curves are upper envelopes to the analytical
curves. Furthermore, we note that the CPT probability difference
$\Delta P_{\mu e}^{\rm CPT}$ is larger for long baseline experiments
with longer baseline lengths and it does not change radically for different
values of the CP violation phase $\delta_{\rm CP}$.

\begin{figure}[th!]
\begin{center}
\includegraphics[height=11cm]{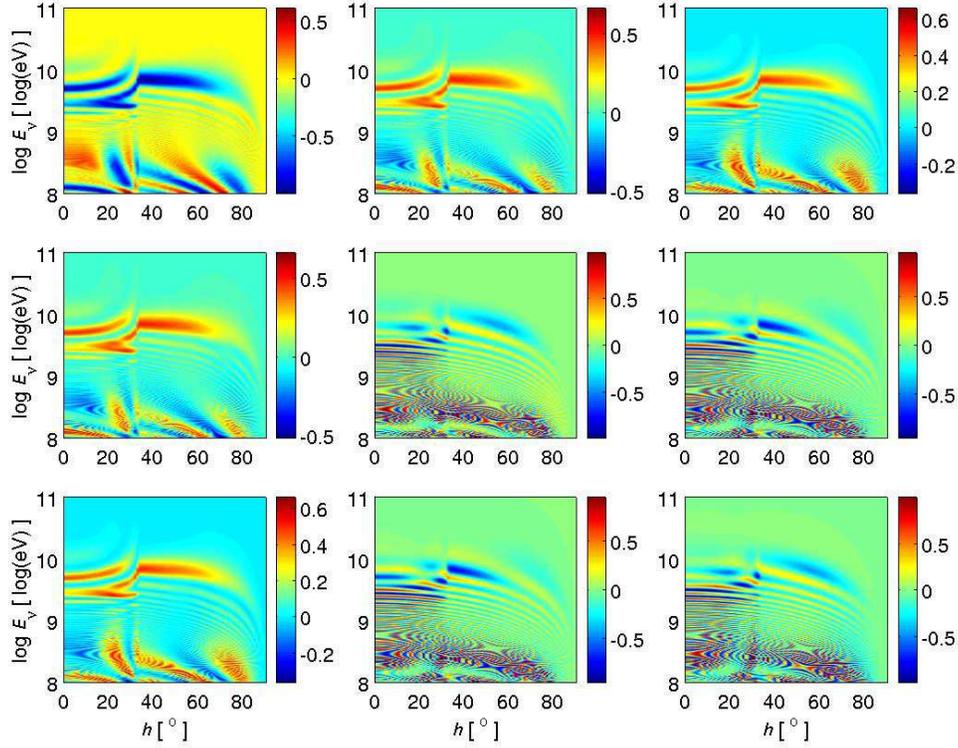}
\end{center}
\caption{\scriptsize The CPT probability differences $\Delta
    P_{\alpha\beta}^{\rm
    CPT}$ ($\alpha,\beta = e,\mu,\tau$) plotted as functions of the
    nadir angle $h$ and the neutrino energy $E_\nu$. The different
    plots show: $\Delta P_{ee}^{\rm CPT}$ (upper-left), $\Delta
    P_{e\mu}^{\rm CPT}$ (upper-middle), $\Delta P_{e\tau}^{\rm CPT}$
    (upper-right), $\Delta P_{\mu e}^{\rm CPT}$ (middle-left), $\Delta
    P_{\mu\mu}^{\rm CPT}$ (middle-middle), $\Delta P_{\mu\tau}^{\rm
    CPT}$ (middle-right), $\Delta P_{\tau e}^{\rm CPT}$ (down-left),
    $\Delta P_{\mu\tau}^{\rm CPT}$ (down-middle), and $\Delta P_{\tau\tau}^{\rm
    CPT}$ (down-right). The fundamental neutrino parameters used are: 
    $\Delta m_{21}^2 = 7.1 \cdot 10^{-5} \, {\rm eV}^2$, $\Delta
    m_{31}^2 = 2.5 \cdot 10^{-3} \, {\rm eV}^2$, $\theta_{12} =
    34^\circ$, $\theta_{13} = 9.2^\circ$, $\theta_{23} = 45^\circ$,
    and $\delta_{\rm CP} = 0$. Furthermore, we have used the
    mantle-core-mantle step-function approximation of the Earth matter
    density profile.}
\label{fig:DeltaPabCPT}
\end{figure}

Finally, in \fig~\ref{fig:DeltaPabCPT}, we present numerical calculations shown
as density plots of the CPT probability differences for neutrinos traversing
the Earth, which are functions of the nadir angle $h$ and the neutrino
energy $E_\nu$.
The numerical calculations are based on the evolution operator method
and Cayley--Hamilton formalism introduced and developed in
\Refs~\cite{Ohlsson:1999xb,Ohlsson:1999um,Ohlsson:2001vp}
and the parameter values used are given in the figure caption. The nadir angle
$h$ is related to the baseline length $L$ as follows. A nadir angle of
$h = 0$ corresponds to a baseline length of $L = 2 R_\oplus$, whereas
$h = 90^\circ$ corresponds to $L = 0$. As $h$ varies from 0 to
$90^\circ$, the baseline length $L$ becomes shorter and shorter. At an
angle larger than $h_0 \equiv \arcsin \tfrac{r}{R_\oplus} \simeq
33.17^\circ$, the baseline no longer traverse the core of the Earth.
The CPT probability differences in \fig~\ref{fig:DeltaPabCPT} might be
of special interest for atmospheric (and to some extent solar)
neutrino oscillation studies, since the plots cover all nadir angle
values and neutrino energies between 100 MeV and 100 GeV. We note from
these plots that for some specific values of the nadir angle and the
neutrino energy the CPT violation effects are rather sizeable.

\section{Summary and Conclusions}
\label{sec:s&c}

In conclusion, we have studied extrinsic CPT violation in three flavor
neutrino oscillations, \ie, CPT violation induced purely by matter in
an intrinsically CPT-conserving context. This has been done by
studying the CPT probability differences for arbitrary matter
density profiles in general and for constant matter density profiles
and to some extent step-function matter density profiles in
particular. We have used an analytical approximation based on first
order perturbation theory and a low-energy approximation derived from
this approximation as well as numerical calculations using the
evolution operator method and Cayley--Hamilton formalism. The different
methods have then been applied to a number of accelerator and reactor
long baseline experiments as well as possible future neutrino factory
setups. In addition, their validity and usefulness have been
discussed. Furthermore, atmospheric and solar neutrinos have been
studied numerically using a step-function matter density profile
approximation to the PREM matter density profile. Our results show
that the extrinsic CPT probability differences can be as large as 5~\%
for certain experiments, but be completely negligible for other
experiments. Moreover, we have found that in general the CPT probability
differences increase with increasing baseline length and decrease with
increasing neutrino energy. All this implies that extrinsic CPT
violation may affect neutrino oscillation experiments in a
significant way. Therefore, we propose to the experimental
collaborations to investigate the effects of extrinsic CPT violation
in their respective experimental setups. However, it seems that for
most neutrino oscillation experiments extrinsic CPT violation effects
can safely be ignored.

Finally, we want to mention that in this paper, we have assumed that
the CPT invariance theorem holds, which means that there will be no
room for intrinsic CPT violation effects in our study, and therefore,
the CPT probability differences will only contain extrinsic CPT
violation effects due to matter effects. However, it has been
suggested in the literature that there might be small intrinsic CPT
violation effects in neutrino oscillations
\cite{Coleman:1998ti,Barger:2000iv}, which might be entangled with the
extrinsic CPT violation effects. The question if such intrinsic and
the extrinsic CPT violation effects could be disentangled from each
other in, for example, realistic long-baseline neutrino oscillation
experiments is still open \cite{Xing:2001ys} and it was not the
purpose of the present study. Actually, this deserves an own complete
systematic study. However, such a study would be highly model
dependent, since intrinsic CPT violation is not present in the
SM. Furthermore, it should be noted that in the above
mentioned references, \Refs~\cite{Coleman:1998ti,Barger:2000iv}, the
intrinsic CPT violation effects were only studied in neutrino
oscillations with two flavors and not with three.

\begin{acknowledgments}
We would like to thank Samoil~M.~Bilenky, Robert Johansson, Hisakazu
Minakata, Gerhart Seidl, H{\aa}kan Snellman, and Walter Winter for
useful discussions and comments.

This work was supported by the Swedish Research Council
(Vetenskapsr{\aa}det), Contract No.~621-2001-1611, 621-2002-3577, the
Magnus Bergvall Foundation (Magn.~Bergvalls Stiftelse), and the
Wenner-Gren Foundations.
\end{acknowledgments}

\appendix

\section{Evolution Operators}
\label{app:evolop}

Neutrino oscillations are governed by the Schr{\"o}dinger equation
[see \eq~(\ref{eq:Schrodinger})]
\begin{equation}
{\rm i} \frac{\d}{\d t} | \nu(t) \rangle = \mathscr{H}(t) | \nu(t) \rangle.
\end{equation}
Inserting $| \nu(t) \rangle = S(t,t_0) | \nu(t_0) \rangle$
[\eq~(\ref{eq:Smatrix})] yields the Schr{\"o}dinger equation for the
evolution operator
\begin{equation}
{\rm i} \frac{\d}{\d t} S(t,t_0) = \mathscr{H}(t) S(t,t_0),
\end{equation}
which we write in flavor basis as
\begin{equation}
{\rm i} \frac{\d}{\d t} S_f(t,t_0) = {\mathscr H}_f(t) S_f(t,t_0).
\label{eq:Sch}
\end{equation}
In what follows, we will assume that the number of neutrino flavors is
equal to three, \ie, $n=3$.
Thus, the total Hamiltonian in flavor basis for neutrinos is given by
\begin{equation}
{\mathscr H}_f(t) = H_f + V_f(t) = U H_m U^\dagger + V_f(t),
\label{eq:Hf}
\end{equation}
where
$$
H_m = \left( \begin{matrix} 0 & 0 & 0 \\ 0 & \delta & 0 \\ 0 & 0 &
\Delta \end{matrix} \right) \quad \mbox{and} \quad V_f(t) = \left(
\begin{matrix} V(t) & 0 & 0 \\ 0 & 0 & 0 \\ 0 & 0 & 0 \end{matrix}
\right)
$$
are the free Hamiltonian in mass basis and the matter potential in
flavor basis, respectively, and $U$ is the leptonic mixing
matrix\footnote{We will use the standard parameterization of the leptonic
mixing matrix \cite{Hagiwara:2002fs}.}.
Here $\delta \equiv \frac{\Delta m_{21}^2}{2E_\nu}$, $\Delta \equiv
\frac{\Delta m_{31}^2}{2E_\nu}$, and $V(t) = \sqrt{2} G_F N_e(t)$ is
the charged-current contribution of electron neutrinos to the matter
potential, where $G_F \simeq 1.16639 \cdot 10^{-23} \, {\rm eV}^{-2}$
is the Fermi weak coupling constant and $N_e(t) = \frac{Y_e}{m_N}
\rho(t)$ is the electron number density with $Y_e$ being average
number of electrons per nucleon (in the Earth: $Y_e \simeq
\tfrac{1}{2}$), $m_N \simeq 939.565330 \, {\rm MeV}$ the nucleon mass, and
$\rho \equiv \rho(t)$ the matter density.
The sign of the matter potential depends on the presence of neutrinos
or antineutrinos. In the case of antineutrinos, one has to change the
sign by the replacement $V(t) \to -V(t)$. Thus, the the total
Hamiltonian in flavor basis for antineutrinos is given by
\begin{equation}
\bar{\mathscr H}_f(t) = H_f - V_f(t) = U H_m U^\dagger - V_f(t).
\label{eq:antiHf}
\end{equation}

Decomposing $U = O_{23} U_{13} O_{12} = O_{23} U'$, we can write the
total Hamiltonian in flavor basis as
\begin{equation}
{\mathscr H}_f(t) = O_{23} \left[ U' \left( \begin{matrix} 0 & 0 &
0 \\ 0 & \delta & 0 \\ 0 & 0 & \Delta \end{matrix} \right) {U'}^\dagger
+ \left( \begin{matrix} V(t) & 0 & 0 \\ 0 & 0 & 0 \\ 0 & 0 & 0
\end{matrix} \right) \right] O_{23}^T \equiv O_{23} H(t) O_{23}^T.
\end{equation}
Here we use the following parameterization for the orthogonal
matrices $O_{23}$ and $O_{12}$ and the unitary matrix $U_{13}$
$$
O_{23} = \left( \begin{matrix} 1 & 0 & 0 \\ 0 & c_{23} & s_{23} \\
0 & -s_{23} & c_{23} \end{matrix} \right), \quad U_{13} = \left(
\begin{matrix} c_{13} & 0 & s_{13} {\rm e}^{-{\rm i}\delta_{\rm CP}}
\\ 0 & 1 & 0 \\ -s_{13} {\rm e}^{{\rm i}\delta_{\rm CP}} & 0 & c_{13}
\end{matrix} \right), \quad O_{12} = \left( \begin{matrix} c_{12} &
s_{12} & 0 \\ -s_{12} & c_{12} & 0 \\ 0 & 0 & 1 \end{matrix} \right),
$$
where $s_{ab} \equiv \sin \theta_{ab}$ and $c_{ab} \equiv \cos
\theta_{ab}$. Here $\theta_{12}$, $\theta_{13}$, and $\theta_{23}$ are
the ordinary vacuum mixing angles and $\delta_{\rm CP}$ is the CP
violation phase. This means that $U$ is given by the standard
parameterization of the leptonic mixing matrix and $U'$ is given by
\begin{equation}
U' = \left( \begin{matrix} c_{13} c_{12} & c_{13} s_{12} & s_{13}
{\rm e}^{-{\rm i} \delta_{\rm CP}} \\ -s_{12} & c_{12} & 0 \\ -s_{13} c_{12}
{\rm e}^{{\rm i} \delta_{\rm CP}} & -s_{13} s_{12} {\rm e}^{{\rm i}
\delta_{\rm CP}} & c_{13} \end{matrix} \right).
\end{equation}

Inserting ${\mathscr H}_f(t) = O_{23} H(t) O_{23}^T$ into the
Schr{\"o}dinger equation, we obtain
\begin{equation}
{\rm i} \frac{\d}{\d t} S(t,t_0) = H(t) S(t,t_0),
\label{eq:SchS}
\end{equation}
where $S(t,t_0) \equiv O_{23}^T S_f(t,t_0) O_{23}$.
Thus, the Hamiltonian $H(t)$ can be written as
\begin{equation}
H(t) = \left( \begin{matrix} c_{13}^2 s_{12}^2 \delta + s_{13}^2
\Delta + V(t) & c_{13} c_{12} s_{12} \delta & c_{13} s_{13} \left(
\Delta - s_{12}^2 \delta \right) {\rm e}^{-{\rm i} \delta_{\rm CP}} \\ c_{13}
c_{12} s_{12} \delta & c_{12}^2 \delta & - s_{13} c_{12} s_{12}
{\rm e}^{-{\rm i} \delta_{\rm CP}} \delta \\ c_{13} s_{13} \left( \Delta -
s_{12}^2 \delta \right) {\rm e}^{{\rm i} \delta_{\rm CP}} & - s_{13} c_{12}
s_{12} {\rm e}^{{\rm i} \delta_{\rm CP}} \delta & s_{13}^2 s_{12}^2 \delta +
c_{13}^2 \Delta \end{matrix}
\right).
\end{equation}
Series expansions of $s_{13}$ and $c_{13}$ when $\theta_{13}$ is
small, \ie, $s_{13} = \theta_{13} + {\mathcal O}(\theta_{13}^3)$
and $c_{13} = 1 + {\mathcal O}(\theta_{13}^2)$, gives up to second order
in $\theta_{13}$
\begin{equation}
H(t) \simeq \left( \begin{matrix} \delta s_{12}^2 + V(t) & \delta
c_{12} s_{12} & \theta_{13} \left( \Delta - \delta s_{12}^2 \right)
{\rm e}^{-{\rm i} \delta_{\rm CP}} \\ \delta c_{12} s_{12} & \delta
c_{12}^2 & - \theta_{13} \delta c_{12} s_{12} {\rm e}^{-{\rm i}
\delta_{\rm CP}} \\ \theta_{13} \left( \Delta - \delta s_{12}^2
\right) {\rm e}^{{\rm i} \delta_{\rm CP}} & - \theta_{13} \delta 
c_{12} s_{12} {\rm e}^{{\rm i} \delta_{\rm CP}} & \Delta \end{matrix} \right).
\end{equation}
Separating $H(t)$ in independent and dependent parts of $\theta_{13}$
yields
\begin{equation}
H(t) = H_0(t) + H', \quad H' = H_1 + H_2,
\label{eq:hh0h1}
\end{equation}
where
\begin{eqnarray}
H_0(t) &=& \left( \begin{matrix} s_{12}^2 \delta + V(t) & c_{12}
s_{12} \delta & 0 \\ c_{12} s_{12} \delta & c_{12}^2 \delta & 0 \\ 0 &
0 & \Delta \end{matrix} \right) \equiv \left( \begin{matrix} h(t)
& \begin{matrix} 0 \\ 0 \end{matrix} \\ \begin{matrix} 0 & 0
\end{matrix} & \Delta \end{matrix} \right), \label{eq:h0t} \\
H_1 &=& \left( \begin{matrix} 0 & 0 & \theta_{13} \left( \Delta -
s_{12}^2 \delta \right) {\rm e}^{-{\rm i} \delta_{\rm CP}} \\ 0 & 0 &
- \theta_{13} c_{12} s_{12} {\rm e}^{-{\rm i} \delta_{\rm CP}} \delta
\\ \theta_{13} \left( \Delta - s_{12}^2 \delta \right) {\rm e}^{{\rm
i} \delta_{\rm CP}} & - \theta_{13} c_{12} s_{12} {\rm e}^{{\rm i}
\delta_{\rm CP}} \delta & 0 \end{matrix} \right) \nonumber\\ &\equiv&
\left( \begin{matrix} 0 & 0 & a \\ 0 & 0 & b \\ a^\ast & b^\ast & 0
\end{matrix} \right), \label{eq:H1} \\ H_2 &=& {\mathcal O}(\theta_{13}^2).
\end{eqnarray}
Here the Hamiltonian $H_1$ is of order $\theta_{13}$, whereas the
Hamiltonian $H_2$ is of order $\theta_{13}^2$.
Note that the Hamiltonian $H'$ is independent of time
$t$. Furthermore, the time-dependent Hamiltonian $H_0(t)$ is only
dependent on the mixing angle $\theta_{12}$.

Inserting \eq~(\ref{eq:hh0h1}) as well as $S(t,t_0) \equiv S_0(t,t_0)
S_1(t,t_0)$
into \eq~(\ref{eq:SchS}) gives
\begin{eqnarray}
&& {\rm i} \left( \frac{\d}{\d t} S_0(t,t_0) \right) S_1(t,t_0) + {\rm i}
S_0(t,t_0) \frac{\d}{\d t} S_1(t,t_0) \nonumber\\
&=& H_0(t) S_0(t,t_0) S_1(t,t_0) + H_1 S_0(t,t_0) S_1(t,t_0).
\end{eqnarray}
Now, assuming that ${\rm i} \frac{\d}{\d t} S_0(t,t_0) = H_0(t) S_0(t,t_0)$
holds implies that we have the equation ${\rm i} \frac{\d}{\d t} S_1(t,t_0) =
H_1(t) S_1(t,t_0)$, where $H_1(t) \equiv S_0^{-1}(t,t_0) H_1 S_0(t,t_0)$,
which can be integrated to give the integral equation
\begin{eqnarray}
S_1(t,t_0) &=& \mathbbm{1} - {\rm i} \int_{t_0}^t H_1(t') S_1(t',t_0)
\d t' \nonumber\\
&=& \mathbbm{1} - {\rm i} \int_{t_0}^t
S_0^{-1}(t',t_0) H_1 S_0(t',t_0) S_1(t',t_0) \d t'.
\end{eqnarray}
Thus, from first order perturbation theory we obtain
\cite{Arafune:1997hd,Akhmedov:2001kd}
\begin{equation}
S(t,t_0) \simeq S_0(t,t_0) - {\rm i} S_0(t,t_0) \int_{t_0}^t
S_0^{-1}(t',t_0) H_1 S_0(t',t_0) \d t'.
\label{eq:1stS}
\end{equation}

Since we assumed before that ${\rm i} \frac{\d}{\d t} S_0(t,t_0) =
H_0(t) S_0(t,t_0)$ holds, we have now to find $S_0(t,t_0)$. We observe
that the $2 \times 2$ submatrix in the upper-left corner of $H_0(t)$ in
\eq~(\ref{eq:h0t}), \ie, $h(t)$, is not traceless. Making
this submatrix traceless yields
\begin{eqnarray}
\tilde{H}_0(t) &=& H_0(t) - \frac{1}{2} {\rm tr \,} h(t) \mathbbm{1}_3
\nonumber\\
&=& \left( \begin{matrix} - \frac{1}{2} (c_{12}^2 - s_{12}^2) \delta +
\frac{1}{2} V(t) & c_{12} s_{12} \delta & 0 \\ c_{12} s_{12} \delta &
\frac{1}{2} (c_{12}^2 - s_{12}^2) \delta - \frac{1}{2} V(t) & 0 \\ 0 &
0 & \Delta - \frac{1}{2} \left[V(t) + \delta\right] \end{matrix}
\right). \nonumber\\ \label{eq:tildeH0t}
\end{eqnarray}
Note that, in general, any term proportional to the identity matrix
$\mathbbm{1}_3$ can be added to or subtracted from the Hamiltonian
$H_0(t)$ without affecting the neutrino oscillation probabilities. In
particular, a term such that the $2 \times 2$ submatrix $h(t)$ in the
upper-left corner of $H_0(t)$ becomes traceless [see
\eq~(\ref{eq:tildeH0t})]. Furthermore, note that the new Hamiltonian
$\tilde{H}_0(t)$ will not be traceless and that the $(3,3)$-element of
$H_0(t)$ will, of course, also be changed by such a transformation.

Instead of solving ${\rm i} \frac{\d}{\d t} S_0(t,t_0) = H_0(t)
S_0(t,t_0)$, we have now to solve ${\rm i} \frac{\d}{\d t} S_0(t,t_0)
= \tilde{H}_0(t) S_0(t,t_0) + \frac{1}{2} {\rm tr \,} h(t)
S_0(t,t_0)$.
The solution to this equation, $S_0(t,t_0)$, has the general form
\cite{Akhmedov:1998ui,Akhmedov:1999uz,Akhmedov:2001kd}
\begin{equation}
S_0(t,t_0) = \left( \begin{matrix} \alpha(t,t_0) & \beta(t,t_0) &
0 \\ -\beta^\ast(t,t_0) & \alpha^\ast(t,t_0) & 0 \\ 0 & 0 & f(t,t_0)
\end{matrix} \right),
\label{eq:S0}
\end{equation}
where the functions $\alpha(t,t_0)$ and $\beta(t,t_0)$ describe the two flavor
neutrino evolution in the $(1,2)$-subsector, in which the $2 \times 2$
submatrix $h$ of $H_0$ acts as the Hamiltonian. In the end of this
appendix, we will derive the analytical expressions for the functions
$\alpha(t,t_0)$ and $\beta(t,t_0)$.
The function $f(t,t_0)$ can, however, immediately be determined to be
\begin{equation}
f(t,t_0) = {\rm e}^{-{\rm i} \int_{t_0}^t \tilde{\Delta}(t') \d t'} \equiv
{\rm e}^{-{\rm i} \Phi(t,t_0)},
\end{equation}
where $\tilde{\Delta}(t) \equiv \Delta - \frac{1}{2} \left[ V(t) + \delta
\right]$ and 
\begin{eqnarray}
\Phi(t,t_0) &\equiv& \int_{t_0}^t \tilde{\Delta}(t') \d t' = \int_{t_0}^t
\left\{ \Delta - \frac{1}{2} \left[ V(t') + \delta \right] \right\} \d
t' \nonumber\\
&=& \left( \Delta - \frac{\delta}{2} \right) \left( t - t_0 \right) -
\frac{1}{2} \int_{t_0}^t V(t') \d t'. \nonumber
\end{eqnarray}

Now, inserting \eqs~(\ref{eq:H1}) and (\ref{eq:S0}) into
\eq~(\ref{eq:1stS}) yields
\begin{equation}
S(t,t_0) \simeq \left( \begin{matrix} \alpha(t,t_0) & \beta(t,t_0)
& -{\rm i} f(t,t_0) A(t,t_0) \\ - \beta^\ast(t,t_0) &
\alpha^\ast(t,t_0) & -{\rm i} f(t,t_0) B(t,t_0) \\
-{\rm i} f(t,t_0) C(t,t_0) & -{\rm i} f(t,t_0) D(t,t_0) &
f(t,t_0) \end{matrix} \right),
\end{equation}
where
\begin{eqnarray}
A(t,t_0) &=& f^\ast(t,t_0) \big\{ a \left[ \alpha(t,t_0)
  I_{\alpha^\ast,t_0}(t,t_0) + \beta(t,t_0) I_{\beta^\ast,t_0}(t,t_0)
  \right]\nonumber\\
 &+& b \left[ \beta(t,t_0) I_{\alpha,t_0}(t,t_0) - \alpha(t,t_0)
  I_{\beta,t_0}(t,t_0) \right] \big\}, \label{eq:Att0} \\
B(t,t_0) &=& f^\ast(t,t_0) \big\{ a \left[ \alpha^\ast(t,t_0)
  I_{\beta^\ast,t_0}(t,t_0) - \beta^\ast(t,t_0) I_{\alpha^\ast,t_0}(t,t_0)
  \right]\nonumber\\
 &+& b \left[ \alpha^\ast(t,t_0) I_{\alpha,t_0}(t,t_0) + \beta^\ast(t,t_0)
  I_{\beta,t_0}(t,t_0) \right] \big\}, \label{eq:Btt0} \\
C(t,t_0) &=& a^\ast I_{\alpha^\ast,t_0}^\ast(t,t_0) - b^\ast
  I_{\beta,t_0}^\ast(t,t_0), \label{eq:Ctt0} \\
D(t,t_0) &=& a^\ast I_{\beta^\ast,t_0}^\ast(t,t_0) + b^\ast
  I_{\alpha,t_0}^\ast(t,t_0) \label{eq:Dtt0}
\end{eqnarray}
with
\begin{equation}
I_{\varphi,t_0}(t,t_0) = \int_{t_0}^t \varphi(t',t_0) f(t',t_0) \d t', \quad
\varphi = \alpha, \alpha^\ast, \beta, \beta^\ast.
\end{equation}
Equations~(\ref{eq:Att0}) and (\ref{eq:Btt0}) can be further
simplified using the following
\begin{eqnarray}
S_0(t_1,t) &=& S_0(t_1,t_0) S_0^\dagger(t,t_0) \nonumber\\
&=& \left( \begin{matrix} \alpha(t_1,t_0) & \beta(t_1,t_0) & 0 \\ -
  \beta^\ast(t_1,t_0) & \alpha^\ast(t_1,t_0) & 0 \\ 0 & 0 & f(t_1,t_0)
\end{matrix} \right) \left( \begin{matrix} \alpha^\ast(t,t_0) & -
  \beta(t,t_0) & 0 \\ \beta^\ast(t,t_0) & \alpha(t,t_0) & 0 \\ 0 & 0 &
  f^\ast(t,t_0) \end{matrix} \right) \nonumber\\
&=& \left( \begin{matrix} \alpha(t_1,t) & \beta(t_1,t) & 0 \\ -
  \beta^\ast(t_1,t) & \alpha^\ast(t_1,t) & 0 \\ 0 & 0 & f(t_1,t)
\end{matrix} \right). \label{eq:S0t1t}
\end{eqnarray}
Considering \eq~(\ref{eq:S0t1t}), one immediately finds that
\begin{eqnarray}
\alpha(t_1,t_0) \alpha^\ast(t,t_0) + \beta(t_1,t_0) \beta^\ast(t,t_0)
&=& \alpha(t_1,t), \label{eq:simpl1}\\
- \alpha(t_1,t_0) \beta(t,t_0) + \beta(t_1,t_0) \alpha(t,t_0) &=&
\beta(t_1,t), \\
- \beta^\ast(t_1,t_0) \alpha^\ast(t,t_0) + \alpha^\ast(t_1,t_0)
\beta^\ast(t,t_0) &=& - \beta^\ast(t_1,t), \\
\alpha^\ast(t_1,t_0) \alpha(t,t_0) + \beta^\ast(t_1,t_0) \beta(t,t_0)
&=& \alpha^\ast(t_1,t), \\
f(t_1,t_0) f^\ast(t,t_0) &=& f(t_1,t). \label{eq:simpl5}
\end{eqnarray}
Thus, using \eqs~(\ref{eq:simpl1}) - (\ref{eq:simpl5}) as well as the
identity $| f(t,t_0) |^2 = f(t,t_0) f^\ast(t,t_0) = 1$, one can write
\eqs~(\ref{eq:Att0}) and (\ref{eq:Btt0}) as
\begin{eqnarray}
A(t,t_0) &=& a I_{\alpha^\ast,t}(t,t_0) - b I_{\beta,t}(t,t_0), \\
B(t,t_0) &=& a I_{\beta^\ast,t}(t,t_0) + b I_{\alpha,t}(t,t_0).
\end{eqnarray}
Now, rotating $S(t,t_0)$ back to the original basis, one finds the
evolution operator for neutrinos in the flavor basis
\begin{eqnarray}
S_f(t,t_0) &=& O_{23}^T S(t,t_0) O_{23} \nonumber\\
&\simeq& \left( \begin{matrix} \alpha & c_{23} \beta -
  {\rm i} s_{23} f A & - s_{23} \beta - {\rm i}
  c_{23} f A \\ - c_{23} \beta^\ast - {\rm i}
  s_{23} f C & S_{22} & S_{23} \\ s_{23}
  \beta^\ast - {\rm i} c_{23} f C & S_{32}
  & S_{33} \end{matrix} \right) \equiv (S_{f,ab}), \nonumber\\
\label{eq:Sf}
\end{eqnarray}
where
\begin{eqnarray}
S_{22} &\equiv& c_{23}^2 \alpha^\ast + s_{23}^2 f - {\rm i} s_{23}
c_{23} f (B + D), \\
S_{23} &\equiv& - s_{23} c_{23} \left( \alpha^\ast - f \right) - {\rm
  i} f \left( c_{23}^2 B - s_{23}^2 D \right), \\
S_{32} &\equiv& - s_{23} c_{23} \left( \alpha^\ast - f \right) + {\rm
  i} f \left( s_{23}^2 B - c_{23}^2 D \right), \\
S_{33} &\equiv& s_{23}^2 \alpha^\ast + c_{23}^2 f + {\rm i} s_{23}
c_{23} f \left( B + D \right)
\end{eqnarray}
with the notation $\alpha \equiv \alpha(t,t_0)$, $\beta \equiv
\beta(t,t_0)$, $f \equiv f(t,t_0)$, $A \equiv A(t,t_0)$, $B \equiv
B(t,t_0)$, $C \equiv C(t,t_0)$, and $D \equiv D(t,t_0)$. 

Similarly, replacing the total Hamiltonian for neutrinos~(\ref{eq:Hf}) with the
total Hamiltonian for antineutrinos~(\ref{eq:antiHf}) in the Schr{\"o}dinger
equation~(\ref{eq:Sch}), the evolution operator for antineutrinos in the flavor
basis becomes
\begin{equation}
\bar S_f(t,t_0) \simeq \left( \begin{matrix} \bar\alpha & c_{23} \bar\beta -
  {\rm i} s_{23} \bar f \bar A & - s_{23} \bar\beta - {\rm i}
  c_{23} \bar f \bar A \\ - c_{23} {\bar\beta}^\ast - {\rm i}
  s_{23} \bar f \bar C & \bar S_{22} & \bar S_{23} \\ s_{23}
  {\bar\beta}^\ast - {\rm i} c_{23} \bar f \bar C & \bar S_{32}
  & \bar S_{33} \end{matrix} \right) \equiv (\bar S_{f,ab}),
\label{eq:antiSf}
\end{equation}
where
\begin{eqnarray}
\bar S_{22} &\equiv& c_{23}^2 \bar\alpha^\ast + s_{23}^2 \bar f - {\rm
  i} s_{23} c_{23} \bar f (\bar B + \bar D), \\
\bar S_{23} &\equiv& - s_{23} c_{23} \left( \bar\alpha^\ast - \bar f
  \right) - {\rm i} \bar f \left( c_{23}^2 \bar B - s_{23}^2 \bar D \right), \\
\bar S_{32} &\equiv& - s_{23} c_{23} \left( \bar\alpha^\ast - \bar f
  \right) + {\rm i} \bar f \left( s_{23}^2 \bar B - c_{23}^2 \bar D \right), \\
\bar S_{33} &\equiv& s_{23}^2 \bar\alpha^\ast + c_{23}^2 \bar f + {\rm
  i} s_{23} c_{23} \bar f \left( \bar B + \bar D \right)
\end{eqnarray}
with the same type of notation as in the neutrino case.

We will now derive the general analytical expressions for the
functions $\alpha(t,t_0)$ and $\beta(t,t_0)$. In order to perform this
derivation, we study the evolution operator in the $(1,2)$-subsector,
which is a separate problem in the rotated basis, and its solution is
independent from the total three flavor neutrino problem. We assume
that the evolution operator in the $(1,2)$-subsector,
$S_{(1,2)}(t,t_0)$, satisfies the Schr{\"o}dinger equation for neutrinos
\begin{equation}
{\rm i} \frac{\d}{\d t}S_{(1,2)}(t,t_0) = h(t) S_{(1,2)}(t,t_0),
\end{equation}
where $h(t)$ is the Hamiltonian and it is given by
\begin{eqnarray}
h(t) &=& \left( \begin{matrix} s_{12}^2 \delta + V(t) & s_{12} c_{12}
  \delta \\ s_{12} c_{12} \delta & c_{12}^2 \delta \end{matrix}
  \right) \nonumber\\
&=& \left( \begin{matrix} - \tfrac{1}{2} \left( c_{12}^2 -
  s_{12}^2 \right) \delta + \tfrac{1}{2} V(t) & s_{12} c_{12} \delta
  \\ s_{12} c_{12} \delta & \tfrac{1}{2} \left( c_{12}^2 - s_{12}^2
  \right) \delta - \tfrac{1}{2} V(t) \end{matrix} \right) +
  \frac{1}{2} \left[ \delta + V(t) \right] \mathbbm{1}_2, \nonumber\\
\end{eqnarray}
see \eqs~(\ref{eq:h0t}) and (\ref{eq:tildeH0t}). Note that the term
proportional to the identity matrix $\mathbbm{1}_2$ in the Hamiltonian
$h(t)$ does not affect neutrino oscillations, since such a term will
only generate a phase factor. Thus, we need not consider this term.
In addition, note that the same term has been subracted from the
Hamiltonian $H_0(t)$ [see \eq~(\ref{eq:tildeH0t})] for the total three
flavor neutrino problem. Thus, it also in this case only gives rise to
a phase factor in the three flavor neutrino evolution operator
$S_0(t,t_0)$ [see \eq~(\ref{eq:S0})], which does not affect the
neutrino oscillations.

The solution to the Schr{\"o}dinger equation in the $(1,2)$-subsector is
\begin{equation}
S_{(1,2)}(t,t_0) = {\rm e}^{-{\rm i} \int_{t_0}^t h(t') \d t'} \equiv
{\rm e}^{-{\rm i} H(t,t_0)},
\end{equation}
where the integrated Hamiltonian, $H(t,t_0)$, is given by
\begin{equation}
H(t,t_0) = \frac{1}{2} \left( \begin{matrix} -\cos 2\theta_{12} \delta
  (t-t_0) + \int_{t_0}^t V(t') \d t' & \sin 2\theta_{12} \delta
  (t-t_0) \\ \sin 2\theta_{12} \delta (t-t_0) & \cos 2\theta_{12}
  \delta (t-t_0) - \int_{t_0}^t V(t') \d t' \end{matrix} \right).
\end{equation}
Since $H(t,t_0)$ is a $2 \times 2$ matrix, the solution can be written
on the following form \cite{Akhmedov:2001kd}
\begin{eqnarray}
S_{(1,2)}(t,t_0) &=& \cos \sqrt{-\det H(t,t_0)} \mathbbm{1}_2
\nonumber\\
&-& {\rm i} \frac{1}{\sqrt{-\det H(t,t_0)}} \sin \sqrt{-\det H(t,t_0)}
H(t,t_0), \label{eq:cossin}
\end{eqnarray}
where the determinant of $H(t,t_0)$, $\det H(t,t_0)$, is given by
\begin{eqnarray}
\det H(t,t_0) &=& -\frac{1}{4} \left[ \cos 2\theta_{12} \delta (t-t_0)
  - \int_{t_0}^t V(t') \d t' \right]^2 - \frac{1}{4} \sin^2
  2\theta_{12} \delta^2 (t-t_0)^2 \nonumber\\
&=& -\frac{1}{4} \bigg[ \delta^2 (t-t_0)^2 - 2 \cos 2\theta_{12}
    \delta (t-t_0) \int_{t_0}^t V(t') \d t' \nonumber\\
&+& \left( \int_{t_0}^t
    V(t') \d t' \right)^2 \bigg].
\end{eqnarray}
Furthermore, the eigenvalues of $H(t,t_0)$ can be found from the
characteristic equation $\det(H(t,t_0) - \Omega \mathbbm{1}_2) = 0$,
which yields $\Omega = \pm \sqrt{-\det H(t,t_0)}$. Note that in
vacuum, \ie, $V(t) = 0 \; \forall t$, it holds that
$\left. \Omega^2 \right|_{V(t) = 0} = \tfrac{1}{4} \delta^2 (t-t_0)^2
\equiv \Omega_{\rm vac}^2$. Now, if one writes the evolution operator
$S_{(1,2)}(t,t_0)$ as
\begin{equation}
S_{(1,2)}(t,t_0) = \left( \begin{matrix} \alpha(t,t_0) & \beta(t,t_0)
  \\ - \beta^\ast(t,t_0) & \alpha^\ast(t,t_0) \end{matrix} \right),
\end{equation}
then, using \eq~(\ref{eq:cossin}), one can identify the functions
$\alpha(t,t_0)$ and $\beta(t,t_0)$. We obtain
\begin{eqnarray}
\alpha(t,t_0) &=& \cos \Omega + {\rm i} \frac{\sin \Omega}{2\Omega}
\left[ \cos 2\theta_{12} \delta (t-t_0) - \int_{t_0}^t V(t') \d t'
  \right], \\
\beta(t,t_0) &=& - {\rm i} \frac{\sin \Omega}{2\Omega} \sin
2\theta_{12} \delta (t-t_0),
\end{eqnarray}
where again
\begin{eqnarray}
\Omega &=& \pm \sqrt{-\det H(t,t_0)} \nonumber\\
&=& \pm \frac{\delta (t-t_0)}{2} \sqrt{ \left[ \cos 2\theta_{12} -
\frac{1}{\delta (t-t_0)} \int_{t_0}^t V(t') \d t' \right]^2 + \sin^2
  2\theta_{12} }.
\label{eq:omega}
\end{eqnarray}

Similarly, for antineutrinos the functions $\bar\alpha(t,t_0)$ and
$\bar\beta(t,t_0)$ become
\begin{eqnarray}
\bar\alpha(t,t_0) &=& \cos \bar\Omega + {\rm i} \frac{\sin
  \bar\Omega}{2\bar\Omega} 
\left[ \cos 2\theta_{12} \delta (t-t_0) + \int_{t_0}^t V(t') \d t'
  \right], \\
\bar\beta(t,t_0) &=& - {\rm i} \frac{\sin \bar\Omega}{2\bar\Omega} \sin
2\theta_{12} \delta (t-t_0),
\end{eqnarray}
which we, in principle, obtain by making the replacement $V(t) \to
-V(t)$ in the expressions for the functions $\alpha(t,t_0)$ and
$\beta(t,t_0)$. Here
\begin{equation}
\bar\Omega = \pm \frac{\delta (t-t_0)}{2} \sqrt{ \left[ \cos 2\theta_{12} +
\frac{1}{\delta (t-t_0)} \int_{t_0}^t V(t') \d t' \right]^2 + \sin^2
2\theta_{12} }.
\label{eq:antiomega}
\end{equation}
Note that $\Omega$ in \eq~(\ref{eq:omega}) and $\bar\Omega$ in
\eq~(\ref{eq:antiomega}) only differ with respect to the sign in front of the
integral of the matter potential.
Thus, from the expressions for $\Omega$ and $\bar\Omega$ we find that
\begin{equation}
\bar\Omega^2 = \Omega^2 + \cos 2\theta_{12} \delta (t-t_0)
\int_{t_0}^t V(t') \d t'.
\end{equation}
Let us now consider some special cases when the relation between
$\Omega$ and $\bar\Omega$ becomes simpler. In the case that
\begin{itemize}
\item $t-t_0 = 0$, one finds $\Omega = \bar\Omega = 0$, which is
  a trivial and non-interesting case.
\item $\delta = 0$, we have degenerated neutrino masses $m_1 = m_2$
  (and negligible solar mass squared difference) or extremely high
  neutrino energy and this leads to $\Omega^2 = \bar\Omega^2 =
  \frac{1}{4} \left[ \int_{t_0}^t V(t') \d t' \right]^2$, which
  implies that $\bar\Omega = \pm \Omega$. Thus, in addition, we have
  $\alpha = \cos \Omega - {\rm i} \frac{\sin\Omega}{2\Omega}
  \int_{t_0}^t V(t') \d t'$, $\bar\alpha = 
  \cos \Omega + {\rm i} \frac{\sin\Omega}{2\Omega} \int_{t_0}^t V(t')
  \d t' = \alpha^\ast$, and $\beta = \bar\beta = 0$.
\item $\cos 2\theta_{12} = 0$ (\eg, $\theta_{12} = 45^\circ$), we have
  maximal mixing in the $(1,2)$-subsector and this leads to $\Omega^2
  = \bar\Omega^2 = \frac{1}{4} \delta^2 (t-t_0)^2 + \frac{1}{4} \left[
  \int_{t_0}^t V(t') \d t' \right]^2$, which again implies that
  $\bar\Omega = \pm \Omega$. In this case, we find $\alpha = \cos
  \Omega - {\rm i} \frac{\sin\Omega}{2\Omega} \int_{t_0}^t V(t') \d
  t'$, $\bar\alpha = \cos \Omega + {\rm i} \frac{\sin\Omega}{2\Omega}
  \int_{t_0}^t V(t') \d t' = \alpha^\ast$, and $\beta = \bar\beta = -
  {\rm i} \frac{\sin\Omega}{2\Omega} \delta (t-t_0)$.
\item $\int_{t_0}^t V(t') \d t' = 0$, one obtains $\Omega^2 =
  \bar\Omega^2 = \frac{1}{4} \delta^2 (t-t_0)^2$, which also implies
  that $\bar\Omega = \pm \Omega$. Furthermore, one has $\alpha =
  \bar\alpha = \cos \Omega + {\rm i} \frac{\sin\Omega}{2\Omega} \cos
  2\theta_{12} \delta (t-t_0)$ and $\beta = \bar\beta = - {\rm i}
  \frac{\sin\Omega}{2\Omega} \sin 2\theta_{12} \delta (t-t_0)$.
\end{itemize}
In addition, if we have close to maximal mixing, \ie, $\theta_{12}
\lesssim 45^\circ$, then we can write $\theta_{12} = \frac{\pi}{4} -
\epsilon$, where $\epsilon$ is a small parameter. Making a series
expansion with the parameter $\epsilon$ as a small expansion
parameter, we obtain
\begin{eqnarray}
\cos 2\theta_{12} &=& 2\epsilon - \frac{4}{3} \epsilon^3 +
\mathcal{O}(\epsilon^5), \\
\bar\Omega &=& \pm \left[ \Omega + \frac{1}{\Omega} \delta (t-t_0)
  \int_{t_0}^t V(t') \d t' \epsilon + \mathcal{O}(\epsilon^2) \right].
\end{eqnarray}

\end{document}